\documentclass[aps,prc,reprint,eqsecnum,showpacs,floatfix]{revtex4-2}

\usepackage{graphicx,epstopdf}
\usepackage{hyperref}
\usepackage{color}
\usepackage{subfigure}
\usepackage{amsmath,mathtools}
\usepackage{amssymb}
\usepackage{epsfig}
\usepackage{amsfonts}
\usepackage{bbm}
\usepackage{bm}
\usepackage{float}
\usepackage{xcolor}
\usepackage{bigints}

\def\bra#1{\bigl\langle{ #1} \bigr|}
\def\ket#1{\bigl|{ #1} \bigr\rangle}

\def\ovlp#1#2{\bigl\langle{ #1}\big|{#2} \bigr\rangle}

 \def\rvec {{\bf r}} \def\pvec {{\bf p}}
\def\hvec {{\bf h}}
\def\qvec {{\bf q}}
\def\kvec {{\bf k}}
\def\hvecF {{\bf h}_{\rm F}}

\def\vec#1{{\bf #1}} % Use bold symbols for vectors
\def\he#1{$^{#1}$He} \def\Tr{{\cal T}r}
\def\LS{\widetildeto{{\bf L}\!\cdot\!{\bf S}}{{\bf L}\!\cdot\!{\bf S}}}
\def\LSp{\widetildeto{{\rm LS}}{{\rm LS}}'}
\def\LSsup{{\rm(LS)}}
\def\LSpup{{\rm(LS')}}
\def\bsigma{{\bm\sigma}}

\def\1{\mathbbm{1}}
\def\EF{e_{\rm F}}
\def\KF{k_{\rm F}}
\def\SF{S_{\rm F}}
\def\S{{\bf S}}
\def\VF{V_{\rm F}}
\def\a0{a_0}
\def\I{{\rm i}}

\def\VLSq{{\tilde V}_{\rm p-h}^{\LSsup}}
\def\VLSr{V_{\rm p-h}^{\LSsup}}

\def\Re{{\cal R}e} \def\Im{{\cal I}m}

\def\ie{{\em i.e.\/}\ }
\def\cf{{\em cf.\/}\ }

\makeatletter
\def\mathcenterto#1#2{\mathclap{\phantom{#1}\mathclap{#2}}\phantom{#1}}
\let\old@widetilde\widetilde
\def\widetildeto#1#2{\mathcenterto{#2}{\old@widetilde{\mathcenterto{#1}{#2\,}}}}
\let\old@widehat\widehat
\def\widehatto#1#2{\mathcenterto{#2}{\old@widehat{\mathcenterto{#1}{#2\,}}}}
\makeatother

\begin{document}

\title{Variational and parquet-diagram calculations for neutron
  matter.\\ IV.  Spin-orbit interactions and linear response}

\author{E.~Krotscheck$^{\dagger\ddagger}$ and J. Wang$^\dagger$}

\affiliation{$^\dagger$Department of Physics, University at Buffalo, SUNY
Buffalo NY 14260}
\affiliation{$^\ddagger$Institut f\"ur Theoretische Physik, Johannes
  Kepler Universit\"at, A 4040 Linz, Austria}

\begin{abstract}
  We develop the parquet-diagram summation method for neutron matter
  interacting via potentials that include spin, tensor, and spin-orbit
  components.  For that purpose, we derive an exact expression for the
  sum of all ring-diagrams in terms effective local particle-hole
  interactions involving the above four operators. The parquet
  equations are closed by deriving the spin-orbit contribution to that
  particle-hole interaction. We show that many-body correlations
  screen the bare spin-orbit potential considerably, and the
  corrections of that screened spin-orbit potential to the other three
  interaction channels are quite small. We apply our method to the
  calculation of the response of neutron matter to density and both
  longitudinal and transverse spin-dependent external fields.
\end{abstract}
\maketitle

\section{Introduction}

\subsection{Interactions: Semi-empirical nucleon-nucleon forces}

Popular models of the nucleon-nucleon forces
\cite{Reid68,Bethe74,Day81,AV18,Wiri84} represent the interaction as a
sum of local functions times correlation operators, \ie
\begin{equation}
\hat v (i,j) = \sum_{\alpha=1}^n v_\alpha(r_{ij})\,
        \hat O_\alpha(i,j),
\label{eq:vop}
\end{equation}
where $r_{ij}=\left|\rvec_i-\rvec_j\right|$ is the distance between
particles $i$ and $j$, and the $\hat O_\alpha(i,j)$ are operators
acting on the spin, isospin, and possibly the relative angular
momentum variables of the individual particles.  According to the
number of operators $n$, the potential model is referred to as a $v_n$
model potential. Semi-realistic models for nuclear matter keep at
least the six base operators and these are
\begin{eqnarray}
\hat O_1(i,j;\hat\rvec_{ij})
        &\equiv& \hat O_c = \1\,,
\nonumber\\
\hat O_3(i,j;\hat\rvec_{ij})
        &\equiv& {\bsigma}_i \cdot {\bsigma}_j\,,
\nonumber\\
\hat O_5(i,j;\hat\rvec_{ij})
&\equiv& S_{ij}(\hat\rvec_{ij})
      \equiv 3({\bm\sigma}_i\cdot \hat\rvec_{ij})
      ({\bsigma}_j\cdot \hat\rvec_{ij})-{\bsigma}_i \cdot {\bsigma}_j\,,
      \nonumber\\
      \hat O_{2n}(i,j;\hat\rvec_{ij}) &=& \hat O_{2n-1}(i,j;\hat\rvec_{ij})
      {\bm\tau}_i\cdot{\bm\tau}_j\,.
\label{eq:operator_v6}
\end{eqnarray}
where $\hat\rvec_{ij} = \rvec_{ij}/r_{ij}$. We will omit the coordinate
arguments when unambiguous.

We have in previous work \cite{v3eos,v3twist} studied $v_6$ models of
the nucleon-nucleon interaction, the next step in the treatment of
successively more realistic nucleon-nucleon interactions is the
inclusion of the spin-orbit interaction. Its operator structure has
the form
\begin{eqnarray}
  \hat O_7(i,j;\rvec_{ij},\pvec_{ij})
  &=&\rvec_{ij}\times\pvec_{ij}\cdot\S\,,
  \nonumber\\
  \hat O_8(i,j;\rvec_{ij},\pvec_{ij})
  &=&\hat O_7(i,j;\rvec_{ij},\pvec_{ij})
       {\bm\tau}_i\cdot{\bm\tau}_j\,,
  \label{eq:operator_v8}
\end{eqnarray}
with $\S\equiv\frac{1}{2}(\bsigma_i+\bsigma_j)$ is the total spin, and
$\pvec_{ij}=\frac{1}{2}(\pvec_i-\pvec_j)$ is the relative momentum
operator of the pair of particles.

This is the subject of the present papers. Unlike the above six
operators, the inclusion of spin-orbit interaction in a variational or
high-order perturbation theory is much more difficult.  Previous
studies used partial-wave expansion of plane waves, because partial
waves are eigenstates of spin-orbit operator. This method becomes
tedious if one goes to higher-order perturbation theory and has been
mostly confined to Bethe-Goldstone calculations
\cite{Spin-OrbitPolls2005}, 3-body coupled cluster \cite{Day81} or
``low-order constraint variational'' (LOCV)
\cite{Tafrihi2015,Tafrihi2017,Tafrihi2018} calculations.

\subsection{Methods: Jastrow-Feenberg Variational and Parquet-Diagrams}

Diagrammatic perturbation theory is one of the standard methods for
dealing with interacting many-body systems. In terms of the paradigms
of that method, it is relatively easy to argue what the {\em
  minimum\/} set of Feynman diagrams should be for a reliable
microscopic treatment of such systems. Let us, for the purpose of
discussion, focus on self-bound systems like \he4, \he3, or nuclear
matter. These are characterized by an {\em equilibrium density\/} with
a negative energy per particle, and {\em saturation\/} meaning that
the energy per particle will eventually begin to increase with
increasing density.

The physical mechanisms behind these effects are clear, as well as the
set of Feynman diagrams that need to be included to deal with these
effects: At {\em high-density\/}, saturation is caused by the
short-ranged interparticle repulsion. That is dealt with by summing
the ladder diagrams, leading to the time-honored Brueckner
or Brueckner-Bethe-Goldstone theory
\cite{Bru54,Bru55,Bru55a,BruecknerLesHouches,BetheGoldstone57,Goldstone57}
which has been the workhorse in nuclear theory for decades.

A second important effect is encountered at low densities: As the
density is lowered to about 2/3 of the saturation density, the system
becomes unstable against infinitesimal density fluctuations; this is
called the ``spinodal point'' where the speed of sound goes to zero.
In a self-bound Fermi system, a second spinodal point appears at very
low density where the interparticle attraction begins to overcome the
Pauli pressure.  Small fluctuations are dealt with in linear response
theory \cite{PinesNoz,ThoulessBook} which implies, in its simplest
version, the calculation of the ring diagrams.

Thus, one is led to the conclusion that the summation of all ring--
and ladder--diagrams of perturbation theory is the least one needs to
do for a consistent description of the equation of state of a
self-bound many-body system over the whole density regime. This set of
diagrams is called the set of ``parquet'' diagrams. Moreover, the
existence of a spinodal point at low density implies that perturbative
calculations encounter a divergence.

While the insight into what is needed is quite obvious, the execution
is far from trivial. A comprehensive treatment of diagrammatic
perturbation theory is found in the seminal paper by Baym and Kadanoff
\cite{BaymKad}, from that it is clear that each two-body vertex depends
on 16 variables (2 incoming and 2 outgoing energy/momentum
sets). Taking energy and momentum conservation as well as isotropy
into account reduces this number to 10 which is still a formidable
task. One must seek approximations, but such steps are normally
ambiguous without further guidance.

A completely different approach was suggested by Jastrow
\cite{Jastrow55} and Feenberg \cite{FeenbergBook}.  For simple,
state-independent interactions as appropriate for electrons or quantum
fluids, the Jastrow-Feenberg ansatz \cite{FeenbergBook} for the wave
function is
\begin{equation}
\Psi_0 = \prod^N_{\genfrac{}{}{0pt}{1}{i,j=1}{i<j}}  f(r_{ij})\Phi_0
\label{eq:Jastrow}
\end{equation}
and its logical generalization to multiparticle correlation functions
has been extremely successful.  Here $\Phi_0$ is a model state
describing the statistics and, when appropriate, the geometry of the
system; for fermions it is normally taken as a Slater determinant but
Bardeen-Cooper-Schrieffer (BCS) states have also been used
\cite{YangClarkBCS,HNCBCS,Fantonipairing,Fabrocinipairing,Fabrocinipairing2,fullbcs,v3bcs}.

One of the reasons for the success of this wave function is that it
provides a reasonable upper bound for the ground state energy
\begin{equation}
E_0 = \frac{\bra{\Psi_0}H\ket{\Psi_0}}{\ovlp{\Psi_0}{\Psi_0}}\,.
  \label{eq:energy}
\end{equation}
For that to work the expectation value \eqref{eq:energy} must, of
course, be calculated with reasonable accuracy. A particularly useful
hierarchy of approximations is the hypernetted-chain summation
technique, it is singled out by the fact that it allows, at every
level of approximation, the unconstrained optimization of the
correlations via the variational principle
\begin{equation}
\frac{\delta E_0}{\delta f}({\bf r}_i,{\bf r}_j) = 0\,.
\label{eq:euler}
\end{equation}
In this case, the method is referred to as the
(Fermi-)Hypernetted-Chain-Euler-Lagrange (F)HNC-EL procedure.

It was quickly realized that the procedure also had exactly the
features of high-density saturation and a low-density spinodal
instability outlined above. For bosons, Sim, Buchler, and Woo
\cite{Woo70} came therefore to the conclusion that ``...it appears
that the optimized Jastrow function is capable of summing all rings
and ladders, and partially all other diagrams, to infinite order.''
This being more a matter of observation than of physical insight,
Jackson, Lande and Smith went back to diagrammatic perturbation theory
and showed in a series of papers \cite{parquet1,parquet2,parquet3}
that indeed the ``optimized HNC equations'' represented an approximate
summation of all parquet diagrams, and determined exactly what these
approximations were. With that, a very practical way to perform
parquet-diagram summations was found. The types of approximations were
singled out by the upper bound property of $E_0$ as the best possible
for the computational price one was willing to pay.  Moreover, while
the derivation of the relevant equations is somewhat complicated, the
resulting equations to be solved numerically were just the
Bethe-Goldstone and the RPA equation that had been solved
individually for decades.

The above holds, in its rigor, only for bosons. A similar systematic
diagrammatic equivalence between the fermion version and parquet-diagram
summations has not been carried out. Rather, the equivalence
has been highlighted as specific sets of diagrams: rings, ladders,
and self-energy insertions \cite{fullbcs}.

The situation is considerably more complicated for realistic
nucleon-nucleon interactions of the form (\ref{eq:vop}). A plausible
generalization of the wave function (\ref{eq:Jastrow}) is the
``symmetrized operator product'' \cite{FantoniSpins,IndianSpins}
\begin{equation}
        \Psi_0^{{\rm SOP}}
        = {\cal S} \Bigl[ \prod^N_{\genfrac{}{}{0pt}{1}{i,j=1}{i<j}} \hat f (i,j)\Bigr] \Phi_0\,,
\label{eq:f_prodwave}
\end{equation}
where
\begin{equation}
  \hat f(i,j) = \sum_{\alpha=1}^n f_\alpha(r_{ij})\,
  \hat O_\alpha(i,j)\,,
  \label{eq:fop}
\end{equation}
and ${\cal S}$ stands for symmetrization. The symmetrization is
necessary because the operators $\hat O_\alpha(i,j)$ and $\hat
O_\beta(i,k)$ do not necessarily commute. We have highlighted recently
\cite{v3twist} (see also Ref. \onlinecite{SpinTwist}) the importance
of a proper symmetrization in cases where the bare interaction is
different in spin-singlet and spin-triplet channels. As an extreme
case, some commutator diagrams would {\em diverge\/} for hard-core
interactions if the correlation functions $f_\alpha(r_{ij})$ were
determined by some simplistic method like LOCV. Due to the
complications of a fully symmetrized variational wave functions, only
very simple implementations were introduced, like omitting commutators
altogether \cite{FantoniSpins} or treating that state-dependent
correlations in a simple chain approximation (single operator chains,
SOC) \cite{IndianSpins}.

Light was shed on the meaning of commutator corrections
again from the point of view of parquet-diagram summations. Smith and
Jackson \cite{SmithSpin} showed for a fictitious system of bosons with
spin, isospin, and tensor forces that the parquet-diagram summation
led to no commutator diagrams, \ie to an optimized Bose-version of
Ref. \onlinecite{FantoniSpins}.

The conclusion is therefore that, unlike originally believed, the
optimized variational wave function \eqref{eq:f_prodwave} contains
{\rm more\/} than just the parquet set; non-parquet diagrams simply
are neglected when commutators are neglected. This does not solve the
problem that these corrections are potentially important. Again,
within na\"ive perturbation theory, the computation of these diagrams
is extremely cumbersome.  We have therefore taken in
Ref. \onlinecite{v3twist} a hybrid approach and included the leading
non-parquet diagrams in a local approximation suggested by the
variational wave function (\ref{eq:f_prodwave}\ref{eq:fop}). As
expected, rather significant changes in the short-ranged structure of
correlations and effective interactions were found.

This paper is devoted to the next step towards including realistic
nucleon-nucleon interactions by dealing with spin-orbit forces.  The
diagrammatic task is somewhat easier than including ``non-parquet''
diagrams because the spin-orbit force acts only in spin-triplet
states. On the other hand it is made much more complicated since
neither the ``chaining'' (as in the summation of ring-diagrams) nor
the parallel connection (as in the summation of the ladder diagrams)
of two spin-orbit operators generates another spin-orbit operator.  We
shall propose very specific approximate schemes that are, again,
inspired by the Jastrow-Feenberg theory.

\section{Variational and Parquet-Diagram Theory for Spin-Orbit Forces}

\subsection{Chain diagrams}
\subsubsection{Chain-diagram summation of spins and tensors}
\label{sssec:chainsCLT}
Let us as an introductory exercise briefly review the summation
of chain diagrams for an interaction of the $V_6$ form \cite{v3eos}.
We assume a {\em local\/} effective particle-hole interaction
\begin{equation}
  \hat V_{\rm p-h}(q) = \sum_{\alpha=1}^6 \tilde V^{(\alpha)}_{\rm p-h}(q)\,
        \hat O_\alpha(i,j),
\label{eq:vphop}
\end{equation}
which is given in momentum space. Note that this is not the bare
interaction, and also {\em not\/} directly related to the $G$-matrix
of Brueckner theory but rather, in the long wavelength limit, to
Landau's Fermi-Liquid Theory \cite{LandauFLP1,LandauFLP2,BaymPethick}
or, at finite wave numbers, to pseudopotentials \cite{ALP78}. In
neutron matter, the operators are projected to the isospin=1 channel
and the above sum goes over the odd values of $\alpha$ only. As usual,
we define the Fourier transforms with a density factor, \ie
\begin{equation}
  \tilde V_{\rm p-h}^{(\alpha)}(q) =\begin{cases}
  \phantom{-}\rho \int d^3r V_{\rm p-h}^{(\alpha)}(r)
  j_0(qr)&\quad\mathrm{for}\quad\alpha = 1\ldots 4,\\
   -\rho \int d^3r V_{\rm p-h}^{(\alpha)}
   (r)j_2(qr)&\quad\mathrm{for}\quad\alpha = 5,6\,.\\
   \end{cases}
\end{equation}
 For further reference we define
\begin{equation}
  \chi_0(\qvec,\hvec;\omega) =
  -\frac{1}{\varepsilon_p-\varepsilon_h-\hbar\omega-\I\eta}-\frac{1}{\varepsilon_p-\varepsilon_h+\hbar\omega+\I\eta}\,.
\end{equation}
Then, the response function of the non-interacting Fermi fluid is the
hole-state average $(1/N)\sum_{\hvec}$ of $
\chi_0(q,\hvec;\omega) $ with the corresponding spin trace
$\Tr_{\bsigma}$.
\begin{equation}
  \chi_0(q;\omega) = \frac{1}{N}\Tr_\bsigma\sum_\hvec\chi_0(q,\hvec;\omega)\,.\label{eq:chi0}
\end{equation}
As a convention, we will label occupied (``hole'') states by $\hvec$,
$\hvec'$, $\hvec_i$ and unoccupied (``particle'') states by $\pvec$,
$\pvec'$, $\pvec_i$; labels $\qvec$, $\qvec'$ and $\qvec_i$ are used
here to represent the particle-hole-transition wave number, \ie, we
always assume $\pvec_i=\hvec_i+\qvec_i$ if there is no ambiguity.

We begin with the chain approximation of the effective interaction
\begin{equation}
  \hat W(\qvec;\omega)=\hat V_{\rm p-h}(\qvec)+\left[\hat V_{\rm p-h}*\chi_0(q;\omega)
  *\hat V_{\rm p-h}\right]+\dots\,,\label{eq:WRPA}
\end{equation}
and the induced interaction
\begin{equation}
  \hat W_{\rm I}(\qvec;\omega)=\hat W(\qvec;\omega) - \hat V_{\rm p-h}(\qvec;\omega)\,.\label{eq:WindRPA}
\end{equation}

Above we have defined the convolution product of two operators
$\hat A \equiv \hat A(\qvec,\hvec,\hvec',\bsigma,\bsigma')$ and
$\hat B\equiv \hat B(\qvec,\hvec,\hvec',\bsigma,\bsigma')$ as
\begin{eqnarray}
  &&\left[\hat A*\chi_0*\hat B\right](\qvec,\hvec,\hvec',\bsigma,\bsigma')
  \nonumber\\
  &=&\frac{1}{N}\Tr_{\bsigma''} \sum_{\hvec''}
  \hat A(\qvec,\hvec,\hvec'',\bsigma,\bsigma'')\chi_0(\qvec,\hvec'';\omega)
  \times\nonumber\\
  &&\qquad\times\hat B(\qvec,\hvec'',\hvec',\bsigma'',\bsigma')\,.
  \label{eq:convol}
\end{eqnarray}

To carry out the summation of chain diagrams, it is convenient to
transform the spin and tensor operators into the longitudinal
and transverse operators  \cite{FrimanNyman,WEISE1977402}
\begin{subequations}
  \label{eq:Qdef}
  \begin{eqnarray}
    \hat  Q_1&\equiv&\1\,\\
  \hat Q_3 &\equiv& \hat L(\hat\qvec) = ({\bsigma}\cdot \hat\qvec)({\bsigma}'\cdot \hat\qvec)\,,\\
  \hat Q_5 &\equiv& \hat T(\hat\qvec) = {\bsigma}\cdot{\bsigma'}-({\bsigma}\cdot \hat\qvec)({\bsigma}'\cdot \hat\qvec)\,.
\end{eqnarray}
\end{subequations}
In the basis $\hat Q_\alpha\in\{\1,\,\hat L,\,\hat
T\}\times\{\1,\tau_1\cdot\tau_2\}$, the convolution product
\eqref{eq:convol} decouples in the individual channels. The effective
interactions \eqref{eq:WRPA} and \eqref{eq:WindRPA} can then be written as
\begin{equation}
  \hat W(\qvec;\omega)=\sum_{\alpha=1}^6\tilde W^{(\alpha)}(q;\omega)\hat Q_\alpha\,,
  \label{eq:WofQ}
\end{equation}
with
\begin{equation}
  \tilde W^{(\alpha)}(q;\omega)=\frac{\tilde V_{\rm p-h}^{(\alpha)}(q)}
  {1-\tilde V_{\rm p-h}^{(\alpha)}(q)\chi_0(q;\omega)}\,.\label{eq:decp}
\end{equation}
and, correspondingly, $\hat W_{\rm
  I}(\qvec;\omega)$.  Thus, the summation of chain diagrams is
straightforward in the first six (or three) operator channels.

\subsubsection{Chain-diagram summation in the spin-orbit channel}
\label{sssec:ChainLS}
The task is more complex for spin-orbit like interactions since the
different channels cannot be decoupled. First, we note that it is an
{\em assumption\/} that the particle-hole interaction can be written
in the simple form $\VLSr(r) {\bf L}\cdot\S$: Even the $G$-matrix
includes, for a bare spin-orbit force, powers of the angular momentum
operator to all orders. That has, so far, precluded the extension of
manifestly microscopic approaches for spin-orbit forces to the parquet
level. This assumption has, nevertheless, been quite popular in
particular in connection with the study of the nuclear response using
Skyrme interactions.  \cite{JPG41_2014,PhysRep536,AnnPhys214,NPA627,NPA658,%
NPA658,PhysRevC.80.024314,PhysRevC.84.059904,PhysRevC.89.044302,PhysRevC.100.064301}.

Since we will frequently need the wave vectors $\hvec$,
$\hvec',\ldots$ in units of the Fermi wave number $\KF$, we also
abbreviate $\hvecF \equiv \hvec/\KF$. Also, let $\Delta
\hvec=\hvec-\hvec'$ and define correspondingly $\Delta\hvecF \equiv
\Delta \hvec/\KF$.  The spatial matrix elements of the spin-orbit
interaction are then
\begin{subequations}
  \begin{align}
   & \bra{\hvec+\qvec, \hvec'}V_{\rm p-h}^{\LSsup}(r){\bf L}\cdot{\bf S}
    \ket{\hvec, \hvec'+\qvec}\nonumber\\
  &=\frac{1}{N}\tilde V_{\rm p-h}^{\rm (LS)}(q)
  \I\hat{\vec{q}}\times\Delta\hvecF\cdot\S
  \equiv\frac{1}{N}\tilde V_{\rm p-h}^{\LSsup}(q)\LS\label{eq:LSphhp} \\
  &\bra{\hvec+\qvec,\hvec'-\qvec}V_{\rm p-h}^{\rm (LS)}(r){\bf L}\cdot{\bf S}
  \ket{\hvec,\hvec'}\nonumber\\
  &=\frac{1}{N}\tilde V_{\rm p-h}^{\LSsup}
  (q)\I\hat\qvec\times\Delta\hvecF\cdot\S
  \equiv\frac{1}{N}\tilde V_{\rm p-h}^{\LSsup}(q)\LS\label{eq:LSpphh}
  \end{align}
  \end{subequations}
with
\begin{eqnarray}
  \tilde V_{\rm p-h}^{\LSsup}(q) &=& \frac{\rho}{2}\int d^3r V_{\rm
    p-h}^{\LSsup}(r) r\KF j_1(qr)\,,\\ \LS &=&
  \I\hat
  \qvec\times\Delta\hvecF\cdot\S\label{eq:LSdef}\,.
\end{eqnarray}
From here on, we shall generally mean the {\em momentum space\/}
representations \eqref{eq:Qdef}, \eqref{eq:LSdef} when we refer to
the operators $\hat O_\alpha$, $\hat Q_\alpha$ or $\LS$. Note in particular
that the operator $\LS$ acts only in spin-space and depends
parametrically on the direction $\hat\qvec$ of momentum transfer
and the difference of the hole wave numbers $\Delta\hvec$.

$\tilde V_{\rm p-h}^{\rm (LS)}(q)$ is defined such that its dimensions
are the same in both coordinate and momentum space.  Note also that we
have defined the operator $\LS$ dimensionless, as a consequence
$\tilde V_{\rm p-h}^{\rm (LS)}(q)$ has the dimension of an energy.

The manipulations to derive an effective interaction including a
spin-orbit potential are rather involved; they will be presented in
appendix \ref{app:vchain} where we will derive a closed-form
expression for the effective interaction \eqref{eq:WRPA}. We focus on
the spin channels of the operator set. The corresponding channels with
isospin $\tau_1\cdot\tau_2$ will follow exactly the same
algebra. Besides the operators $\hat Q_1$, $\hat Q_3$ and $\hat Q_5$,
we need to define three additional operators
\begin{subequations}
\begin{eqnarray}
  \hat Q_7 &\equiv& \left[(\hat\qvec\times\hvecF)\cdot\bsigma\right]
  \left[(\hat\qvec\times\hvecF')\cdot\bsigma'\right]\,,\\
  \hat Q_9 &\equiv& 2(\hat\qvec\times\hvecF)\cdot(\hat\qvec\times\hvecF')\,.
\label{eq:Q79def}
\end{eqnarray}
and
\begin{equation}
\LSp \equiv \frac{\I}{2}\left[(\hat\qvec\times\hvecF)\cdot\bsigma -
  (\hat\qvec\times\hvecF')\cdot\bsigma'\right]\,.\label{eq:LSpdef}
\end{equation}
  \end{subequations}

The set of operators $\{\hat Q_\alpha\}$ has the properties
\begin{equation}
  \left[\hat Q_\alpha*\chi_0*\hat Q_\beta\right] =
  \begin{cases}\phantom{\frac{1}{2}}\chi_0(q;\omega)
    \hat Q_\alpha\delta_{\alpha\beta} &\ \text{for}\ \alpha =
    1,3,5,\\ \phantom{\frac{1}{2}}\chi_0^{(\perp)}(q;\omega)\hat
    Q_\alpha\delta_{\alpha\beta}&\ \text{for}\ \alpha =
    7,9
  \end{cases}\label{eq:Qortho}
\end{equation}
with the ``transverse'' Lindhard function
\begin{equation}
  \chi_0^{(\perp)}(q;\omega)=\frac{1}{N}\Tr_\bsigma\sum_\hvec|\hat \qvec\times\hvecF|^2\chi_0(\qvec,\hvec;\omega)\,.\label{eq:chi0trans}
\end{equation}
The explicit form of $\chi_0^{(\perp)}(q;\omega)$ is given in Appendix
\ref{app:chi0trans}.

In terms of these operators, the effective interaction
consists of two contributions: One that contains only even powers
of the spin-orbit interaction and one that contains odd powers.
For a compact representation, we define {\em energy-dependent\/}
``particle-hole'' interactions in the central and the transverse channel
\begin{subequations}
\begin{eqnarray}
  \tilde V_{\rm p-h}^{\rm (c)}(q;\omega)&\equiv& \tilde V_{\rm p-h}^{\rm (c)}(q)+
  \frac{1}{4}\chi_0^{(\perp)}(q;\omega)\left[\VLSq(q)\right]^2\nonumber\\
  &&\,,\label{eq:Vcredef}\\
  \tilde V_{\rm p-h}^{\rm (T)}(q;\omega)&\equiv&
  \tilde V_{\rm p-h}^{\rm (T)}(q)+
  \frac{1}{8}\chi_0^{(\perp)}(q;\omega)\left[\VLSq(q)\right]^2\nonumber\\
  &&\,.\label{eq:VTredef}
\end{eqnarray}
\end{subequations}
There is no energy-dependent correction to the longitudinal channel,
we nevertheless define, for a symmetric notation, $\tilde V_{\rm
  p-h}^{\rm (L)}(q;\omega)\equiv \tilde V_{\rm p-h}^{\rm (L)}(q)$. The
effective interactions for $\alpha=1\,\dots 3$ can contain only even
numbers of spin-orbit operators, we can write it as

\begin{equation}
  \hat W^{(\rm even)}(\qvec,\hvec,\hvec',\bsigma,\bsigma';\omega) =
  \sum_{\substack{\alpha\,{\rm odd}}}^9 \tilde W^{(\alpha)}(q;\omega)\hat Q_\alpha
\end{equation}
with
\begin{equation}
  \tilde W^{(\alpha)}(q;\omega)
  = \frac{\tilde V_{\rm p-h}^{(\alpha)}(q;\omega)}
    {1-\chi_0(q;\omega)\tilde V_{\rm p-h}^{(\alpha)}(q;\omega)}
\end{equation}
for $\alpha=1,\ldots, 5$ and
  \begin{eqnarray}
    \tilde W^{(7)}(q;\omega) &=&\frac{1}{4}\frac{\left[\VLSq(q)\right]^2\chi_0(q;\omega)}{1-\chi_0(q;\omega)\tilde V_{\rm p-h}^{\rm (c)}(q;\omega)}\,,\\
    \tilde W^{(9)}(q;\omega) &=&\frac{1}{8}\frac{\left[\VLSq(q)\right]^2\chi_0(q;\omega)}
   {1-\chi_0(q;\omega)\tilde V_{\rm p-h}^{\rm (T)}(q;\omega)}\,.
   \end{eqnarray}
  For odd numbers of spin-orbit operators we get
\begin{align}
  &\hat W_{\rm LS}^{(\rm odd)}(\qvec,\hvec,\hvec',\bsigma,\bsigma';\omega)
  \nonumber\\
  &=
  \frac{\VLSq(q)}{1-\chi_0(q;\omega)\tilde V_{\rm p-h}^{\rm (c)}(q;\omega)}
  \left[\LS - \LSp\right]\nonumber\\
  &+\frac{\VLSq(q)}{1-\chi_0(q;\omega)\tilde V_{\rm p-h}^{\rm (T)}(q;\omega)}
  \LSp\nonumber\\
  &\equiv\tilde W_{\rm LS}^{\LSsup}(q;\omega)\LS  +
  \tilde W_{\rm LS}^{\LSpup}(q;\omega)\LSp\,.
 \label{eq:VLSodd}
\end{align}

\subsubsection{Response functions}

The response functions describe the response of a system to an
external perturbation. In our case, the external fields may be a scalar
field or a longitudinal and transverse spin-dependent field
\cite{ALBERICO1982429,Pines:1988gik}
\begin{equation}
  h^{\rm (c)}_{\rm ext}(\qvec), \qquad h^{\rm (L)}_{\rm ext}(\qvec)\hat\qvec\cdot\bsigma,
  \quad\text{and}\quad h^{\rm (T)}_{\rm ext}(\qvec)(\hat\qvec\times\bsigma)
  \label{eq:extfields}\,.
\end{equation}

The simplest approximation to the (spin--)density response function
$\chi(q;\omega)$ is the ``random-phase approximation''
(RPA) \cite{PiB51,PiB52a} which implies the summation of chain
diagrams as carried out above. One can beyond that by including
two-particle-two-hole states in the excitation operator
\cite{PhysRev.126.2231,SRPA83,SRPA87,Wambach88,PhysRevC.81.024317},
the procedure is known as ``second RPA (SRPA)'' in nuclear
physics. When built upon a correlated ground state, the method has
been termed ``Dynamic Many-Body Theory (DMBT)'' and has led to
unprecedented agreement between experiments and theories for the helium
liquids \cite{2p2h,Nature_2p2h,eomIII,skw4lett,2dhe4}. These
complications are expected to be much less important in neutron matter
due to its lower density.

In the RPA, the response to the three external field
\eqref{eq:extfields} ({\em plus\/} possible isospin components) is
\begin{eqnarray}
  &&\chi_\alpha(q,\omega) = \frac{\chi_0(q,\omega)}
      {1-\tilde V_{\rm p-h}^{(\alpha)}(q,\omega)\chi_0(q,\omega)}\nonumber\\
      &=& \chi_0(q,\omega) +
      \chi_0(q,\omega)\tilde W^{(\alpha)}(q,\omega)\chi_0(q,\omega)
\label{eq:chialpha}
\end{eqnarray}
for $\alpha=1\dots6$ where the $\tilde V_{\rm p-h}^{(\alpha)}(q,\omega)$
contain, in our case, the energy-dependent corrections
\eqref{eq:Vcredef} and \eqref{eq:VTredef} originating
from the spin-orbit potential.

We are not aware of external fields that couple directly to the
hole momenta and thus would probe directly the components
$\tilde W^{(7)}(q,\omega)$, $\tilde W^{(9)}(q,\omega)$, and the
spin-orbit components of the effective interaction.

\subsubsection{Local approximation}
\label{sssec:LocalChains}

The derivation was so far exact in the sense that we have only assumed
that the particle-hole irreducible interaction $\hat V_{\rm p-h}(\qvec)$
can be written in the form of a $v_8$ potential.  As we can see, the
total effective interaction can {\em not\/} be represented in a
$v_8$ form, but we were able to derive a closed form in terms of a small
number of operators. The exact form derived above would certainly be
useful for the study of $P$-wave pairing in neutron matter where a
very accurate representation of the effective interaction at the Fermi
surface is needed.

This is not the purpose of the present work, rather we seek an optimal
approximation of the interaction in a $v_8$ form for the purpose of a
complete summation of the parquet class of diagrams.  Following and
generalizing the procedure defined in the parquet-diagram papers
\cite{parquet1,parquet2,SmithSpin,v3eos}, we define the local
approximations for all quantities by calculating the hole-state
average
\begin{equation}
  \tilde W^{(\alpha)}(q)=\frac{
  \Tr_{\bsigma\bsigma'}\displaystyle\int \frac{d^3h d^3h'}{\VF^2}
  \hat W(\qvec,\hvec,\hvec')\hat Q_\alpha(\qvec)}
  {\Tr_{\bsigma\bsigma'}\left\langle\hat Q^2_\alpha\right\rangle}
    \,.\label{eq:intrace}
\end{equation}
where $\VF = 4\pi\KF^3/3$ is the volume of the Fermi sphere, and it is
assumed that the effective interaction is written in the form
\eqref{eq:WofQ}.  For the first six operators, the hole-state
integration is trivial, and Eq. \eqref{eq:intrace} reduces to
\begin{equation}
  \tilde W^{(\alpha)}(q) = \frac{\Tr_{\bsigma\bsigma'}
  \hat O_\alpha(\hat\qvec) \hat W(\qvec,\hvec,\hvec')}{\Tr _{\bsigma\bsigma'}\hat O_\alpha^2(\hat\qvec)}\,.
\label{eq:traces}
\end{equation}
The spin-orbit term can be done similarly: For a local interaction
in $v_8$ form we get the
\begin{eqnarray}
  &&\frac{1}{\nu^2} \Tr_{\bsigma\bsigma'}\int \frac{d^3h d^3h'}{\VF^2}
    \hat W(\qvec)
  \left(-\I\hat\qvec\times\Delta \hvecF\right)\cdot\S\nonumber\\
  &=&\frac{1}{\nu^2} \tilde W^{\LSsup}(q) \Tr_{\bsigma\bsigma'}\int \frac{d^3h d^3h'}{\VF^2}
  \left[
    (\hat\qvec\times\Delta\hvecF)\cdot\S\right]^2
  \nonumber\\
  &=&  \frac{1}{2}\tilde W^{\LSsup}(q)\int \frac{d^3h d^3h'}{\VF^2}
  \left|\hat\qvec\times\Delta\hvecF\right|^2
  =\frac{2}{5}\tilde W^{\LSsup}(q)\,,\nonumber\\
  &&\label{eq:LStrace}
\end{eqnarray}
where $\nu$ is the degree of degeneracy of single-particle states.

With that we can derive a local approximation for the full effective
interaction \eqref{eq:decp} in the $v_8$ (or $v_4$) form. The trace
\eqref{eq:LStrace} with the components $\tilde W^{(7)}(q;\omega)$
and $\tilde W^{(9)}(q;\omega)$ is zero. The only non-trivial quantity
is $\tilde W^{\LSpup}(q;\omega)$ for which we obtain
\begin{widetext}
\begin{eqnarray}
 && \frac{5}{2\nu^2}
 \Tr_{\bsigma\bsigma'}\int \frac{d^3h d^3h'}{\VF^2}\tilde W^{\LSpup}(q;\omega)\LSp
 \left[-\I\hat\qvec\times(\hvecF-\hvecF')\right]\cdot\S
 \nonumber\\
 &=&
   \frac{5}{8\nu^2} \tilde W^{\LSpup}(q;\omega)
   \Tr_{\bsigma\bsigma'}\int \frac{d^3h d^3h'}{\VF^2}
   \left[(\hat\qvec\times\hvecF)\cdot\bsigma
     -(\hat\qvec\times\hvecF')\cdot\bsigma'\right]
 \biggl[\hat\qvec\times(\hvecF-\hvecF')\cdot(\bsigma+\bsigma')\biggr]\nonumber\\
   &=&\frac{1}{2}\tilde W^{\LSpup}(q;\omega)\label{eq:WLSfull}
\end{eqnarray}
\end{widetext}
such that the localized effective spin-orbit interaction is simply

\begin{eqnarray}
  \tilde W^{\LSsup}_{\rm local}(q;\omega)
  &=& \tilde W^{\LSsup}(q,\omega)+\frac{1}{2}\tilde W^{\LSpup}(q,\omega)
  \nonumber\\
  &=& \frac{1}{2}\frac{\VLSq(q)}{
    1-\chi_0(q;\omega)\tilde V_{\rm p-h}^{\rm (c)}(q;\omega)}\nonumber\\
    &+&\frac{1}{2}\frac{\VLSq(q)}{
    1-\chi_0(q;\omega)\tilde V_{\rm p-h}^{\rm (T)}(q;\omega)}
      \,.\label{eq:VLSlocal}
  \end{eqnarray}
With that, we have defined a reasonably compact form of the effective
interaction $\hat W_{\rm local}(\qvec;\omega)$ in the $v_8$ form. Since we
will not use the non-local version in our further considerations we
shall henceforth omit the subscript ``local''.

The practical implementation of a full parquet-level calculation
requires yet another step. We have above defined a local, but energy-dependent 
effective interaction. To define a static effective
interaction, we follow the procedure outlined in
Refs. \onlinecite{parquet1} and \onlinecite{parquet2} and generalized
to fermions in Ref. \onlinecite{fullbcs,v3eos}:

\begin{enumerate}
\item{} Assume a static effective particle-hole
  interaction $\hat V_{\rm p-h}(\qvec)$. Calculate the density--density response
  function which we can write as
  \begin{equation}
    \chi_\alpha(q,\omega) = \chi_0(q,\omega)
    + \chi_0(q,\omega)\tilde W^{(\alpha)}(q,\omega) \chi_0(q,\omega)
    \end{equation}
  and, from that, the static structure functions $S_\alpha (q)$
  \begin{equation}
    S_\alpha(q) = -\int \frac{d\omega}{\pi}\Im \chi_\alpha(q,\omega)
  \end{equation}
\item{}
  Define a {\em static\/} effective interaction such that the second-order
  response function
  \begin{equation}
  \chi_\alpha(q,\omega) = \chi_0(q,\omega)
    + \chi_0(q,\omega)\tilde W^{(\alpha)}(q) \chi_0(q,\omega)
  \end{equation}
  leads to the same $S(q)$. Combining the two relationships
  we can define an energy independent effective interaction as
  \begin{equation}
    \tilde W^{(\alpha)}(q)=\frac{\bigintssss \frac{\displaystyle d\omega}
      {\displaystyle\pi}
      \Im \chi_0(q,\omega)\tilde W^{(\alpha)}(q,\omega) \chi_0(q,\omega)}
    {\bigintssss \displaystyle \frac{\displaystyle d\omega}{\displaystyle\pi}
      \Im \chi_0^2(q,\omega)}\,.\label{eq:localization}
    \end{equation}
\end{enumerate}
We use the same procedure to define an energy-independent interaction
in the spin-orbit channel. We stress, of course, that the dynamic
interaction $\hat W(\qvec,\omega)$ is replaced by the static $\hat
W(\qvec)$ only for the ground state correlations. In both the calculation
of the dynamic structure function to be discussed below, as well as in
the calculation of pairing phenomena \cite{fullbcs,v3bcs} we keep the
full energy dependence.

\subsection{Particle-hole potential with spin-orbit interactions}
\label{ssec:Vph}

For bosons, the situation is very simple: One ends up, up to an
undetermined function $V_{\rm I}(r)$, with exactly the HNC-EL
equations or the ``localized parquet'' equations
\cite{parquet2,parquet3}. The correction $V_{\rm I}(r)$ is identified,
in Jastrow-Feenberg theory, with the contribution from ``elementary
diagrams'' and multiparticle correlations. In the language of parquet
theory it stems from totally irreducible diagrams
\cite{TripletParquet}.

The situation is somewhat more complicated for fermions due to the
multitude of exchange contributions. The RPA form
\eqref{eq:WRPA} specifies that one keeps only the simplest exchange
loops. The relevant approximations have been dubbed as FHNC//0 or
parquet//0. More complicated exchange diagrams are also important and
routinely kept \cite{v3eos,v3twist} but we shall restrict ourselves
for the purpose of discussions to the simplest case. A brief
discussion of how we include exchange effects will be given in Sec.
\ref{ssec:exchange}.

We take here a similarly pragmatic approach as what is behind
Jastrow-Feenberg theory: Rather than calculating some low-order
contributions to high precision and then getting stuck doing
high-order terms, we use local approximations of the kind suggested by
the Jastrow-Feenberg wave function for those terms that are otherwise
omitted.
In this case, the effective interaction $\hat V_{\rm
  p-h}(\qvec)$ is structurally identical to that for bosons, \ie,
\begin{eqnarray}
  V_{\rm p-h}(r) &=& \frac{\hbar^2}{m}\left|\nabla\sqrt{1+\Gamma_{\!\rm dd}(r)}\right|^2\nonumber\\
  &+& [1+\Gamma_{\!\rm dd}(r)]v(r)
  +\Gamma_{\!\rm dd}(r)W_{\rm I}(r)\label{eq:Vph}\,.
\end{eqnarray}
Here, $\Gamma_{\!\rm dd}(r)$ is the direct correlation
function.
This function is rigorously defined as the sum of all
``direct-direct'' diagrams of the FHNC diagrammatic summation method
\cite{JohnReview,polish}. In the simplest FHNC//0 or parquet//0
approximation, it is related to the density static structure function by
\begin{equation}
  \tilde\Gamma_{\!\rm dd}(q) = \frac{\SF(q)-S(q)}{\SF^2(q)}\,.
\end{equation}
The induced interaction $W_{\rm I}(r;\omega)$ defined in
Eq. \eqref{eq:WindRPA} is approximated by a local function $W_{\rm
  I}(r)$ by the same procedure that was used to define an 
energy-independent effective interaction, \ie we simply have
\begin{equation}
  W_{\rm I}(r) = W(r)- V_{\rm p-h}(r)\,.
  \label{eq:Windlocal}
\end{equation}
Beginning with $v_6$ interactions, the correlation functions
become angular-dependent. The expression \eqref{eq:Vph}
\begin{eqnarray}
  \hat V_{\rm p-h}(\rvec) &=&
  \frac{\hbar}{m}\left|\nabla\psi(\rvec)\right|^2\label{eq:Vphparquet}\\
  &+& \psi^*(\rvec)\left[\hat V(r)+\hat V_I(\rvec) + \hat W_I(\rvec)\right]\psi(\rvec)
  - \hat W_I(\rvec)\nonumber
\end{eqnarray}
where we have omitted all spin (and possibly isospin) dependencies for
ease of writing. The $\hat V_I(\rvec)$ is in our case the important
contribution of ``twisted-chain'' diagrams \cite{v3twist}.

The pair wave functions $\psi(\rvec)$ are related to
the direct correlation functions by
\begin{equation}
  \left|\psi(\rvec)\right|^2 = 1+\Gamma_{\!\rm dd}(r)
\end{equation}
The short--ranged structure of the pair wave function $\psi(\rvec)$ is
determined by a Bethe-Goldstone like equation \cite{fullbcs}
\begin{eqnarray}
  &&\psi^*(\rvec)\left[-\frac{\hbar^2}{2m}\nabla^2 + \hat v(\rvec) +\hat V_I(\rvec)
    + \hat W_{\rm I}(\rvec)\right]
  \psi(\rvec)\nonumber\\
  &=&\left[t(k)(1-\SF^{-1}(k))\hat\Gamma_{\!\rm dd}(k)\right]^{\cal F}(r)\,.
  \label{eq:ELSchr}
\end{eqnarray}
Alternatively
\cite{ScottMozkowski,OBI76,Tafrihi2015,Tafrihi2017,Tafrihi2018} the
short-ranged structure has been determined by an effective
Schr\"odinger equation supplemented by a ``healing'' condition (LOCV
method).

In the case of $v_6$ model interactions, the products of wave
functions and potentials decouple in the projector channels
$\hat P_\alpha\in\{\hat P_s, \hat P_{t+}, \hat P_{t-}\}$.  The only new
aspect is that we must keep the angular dependence of the tensor
correlations in the kinetic energy term
$\left|\nabla\psi(\rvec)\right|^2$ in Eq. \eqref{eq:Vphparquet}. We
have described this in Ref. \onlinecite{v3eos}, see also
Ref. \onlinecite{SmithSpin}.

The situation becomes again more complicated if spin-orbit components
are included. There are two origins of a spin-orbit contributions
to the particle-hole interaction: One comes from the terms
\[\psi^*(\rvec) v_{\rm LS}(r){\bf L}\!\cdot\!{\bf S}\psi(\rvec)\]
  and
  \[\psi^*(\rvec)W_I^{\LSsup}(r){\bf L}\!\cdot\!{\bf S}
  \psi(\rvec)-W_I^{\LSsup}(r){\bf L}\!\cdot\!{\bf S}\,\]
The spin-orbit operator commutes with a spherically symmetric wave
function, hence we can write this term as
\begin{equation}
  V_{\rm p-h}^{\LSsup}(r) =
  \left[v_{\rm LS}(r)(1+\Gamma_{\rm dd}(r))+W_{\rm I}^{\LSsup}(r)\Gamma_{\rm dd}(r)\right]
  \,.\label{eq:VphLS}
\end{equation}

A second contribution to the particle-hole interaction in spin-orbit
channel comes from the modification of the wave function due to a
spin-orbit term in Eq. \eqref{eq:ELSchr}. That correction goes to zero
for both $r\rightarrow 0$ and $r\rightarrow \infty$
\cite{OBI76,Tafrihi2015}.  Considering the fact that, as we shall see,
the spin-orbit corrections are small anyway, we follow here the
strategy of Ref. \onlinecite{OBI76} and disregard this term. As
already mentioned above, we can also safely disregard the
``twisted-chain'' corrections \cite{v3twist}.  These corrections are
important when the interactions in the spin-singlet and the
spin-triplet case are very different as it is the case for both the
Reid \cite{Reid68} and the Argonne \cite{AV18} potentials. The
spin-orbit force acts only in spin-triplet channels, one does
therefore not expect significant corrections from non-parquet
contributions.
\begin{widetext}
\subsection{Energy Correction}

Let us first look at the direct part of the
second-order correction to the energy
\begin{eqnarray}
\Delta E_2 &=&-\frac{N^3}{2}\sum_{\sigma_p\sigma_{p'}\sigma_h\sigma_{h'}}
 \int \frac{d^3q d^3h d^3h'}{(2\pi)^9\rho^3}
 \frac{\left|\bra{\hvec+\qvec\,\sigma_p,\hvec'-\qvec\sigma_{p'}}\hat v
   \ket{\hvec\sigma_h,\hvec'\sigma_{h'}}\right|^2}{e_p+e_{p'}-e_h-e_{h'}}
\end{eqnarray}
Focusing on the spin-orbit contribution we get from
\eqref{eq:LSpphh}
\begin{equation}
  \bra{\hvec+\qvec\,\sigma_p,\hvec'-\qvec\,\sigma_{p'}}
  v_{\rm LS}(r){\bf L}\cdot{\bf S}\ket{\hvec\,\sigma_h,\hvec'\sigma_{h'}}
  = \frac{1}{N}
  \tilde V_{\rm LS}(q)  \bra{\sigma_p\sigma_{p'}}\LS\ket{\sigma_h\sigma_{h'}}
\end{equation}
and for the energy correction
\begin{equation}
\frac{\Delta E_2}{N}
 =-\frac{1}{2}
\int \frac{d^3q}{(2\pi)^3\rho} \tilde v_{\rm LS}^2(q)
\int \frac{d^3h d^3h'}{(2\pi)^6\rho^2}
\sum_{\sigma_p\sigma_{p'}\sigma_h\sigma_{h'}}\frac{
  \left|\bra{\sigma_p\sigma_p'}\LS\ket{\sigma_h\sigma_h'}\right|^2}
{e_p+e_{p'}-e_h-e_{h'}}
\end{equation}
The matrix element of the spin-orbit operators evaluate to
\begin{equation}
  \sum_{\sigma_p\sigma_{p'}\sigma_h\sigma_{h'}}\left|\bra{\sigma_p\sigma_p'}\LS\ket{\sigma_h\sigma_h'}\right|^2
  = \Tr_{\sigma_p\sigma_{p'}}(\hat\qvec\times\Delta\hvecF\cdot\S)^2\nonumber\\
  =\frac{\nu^2}{2}\left|\hat\qvec\times\Delta\hvecF\right|^2
\end{equation}
such that
\begin{equation}
\frac{\Delta E_2}{N}
 =-\frac{1}{4}
\int \frac{d^3q}{(2\pi)^3\rho} \tilde v_{\rm LS}^2(q)
\int \frac{d^3h d^3h'}{\VF^2}\frac{\left|\hat\qvec\times\Delta\hvecF\right|^2}
 {e_p+e_{p'}-e_h-e_{h'}}\,.
\end{equation}
To go higher orders, use \cite{FetterWalecka}
\begin{equation}
  \int_0^\infty \frac{d\omega}{\pi}\Im \chi_0(\qvec,\hvec;\omega)
  \chi_0(-\qvec,\hvec';\omega)
  = \frac{1}{e_p+e_{p'}-e_h-e_{h'}}
\end{equation}
In our specific case we get
\begin{eqnarray}
  \int \frac{d^3h d^3h'}{V_{\rm F}^2}
  \frac{\left|\hat\qvec\times\Delta\hvecF\right|^2}{e_p+e_{p'}-e_h-e_{h'}}
  &=&2\int_0^\infty \frac{d\omega}{\pi}\Im
  \int \frac{d^3h d^3h'}{V_{\rm F}^2}
  \left|\hat\qvec\times\Delta\hvecF\right|^2 \chi_0(\qvec,\hvec;\omega)
  \chi_0(-\qvec,\hvec';\omega)\nonumber\\
  &=&4\int_0^\infty \frac{d\omega}{\pi}\Im\chi_0(q;\omega)
  \chi_0^{(\perp)}(q,\omega)
\end{eqnarray}
and with that
\begin{eqnarray}
\frac{\Delta E_2}{N}
 &=&-\frac{1}{2}
\int \frac{d^3q}{(2\pi)^3\rho}\tilde v_{\rm LS}^2(q)\int_0^\infty\frac{d\omega}{\pi}
\Im\chi_0(q;\omega)\chi_0^{(\perp)}(q,\omega)\,.\label{eq:DE2LS}
\end{eqnarray}
This was so far standard second-order perturbation theory. The
expression \eqref{eq:DE2LS} is now easily generalized to the sum of
all ring-diagrams by replacing one of the bare interactions $\tilde
v_{\rm LS}(q)$ by $\tilde V_{\rm p-h}^{\LSsup}(q)$, one by $\tilde
W^{\LSsup}(q)$ defined in Eq. \eqref{eq:VLSlocal} and performing the
usual coupling constant integration \cite{rings}.  We hasten to point
out that using the expression \eqref{eq:VLSodd} leads to the same
answer.
\begin{eqnarray}
\frac{\Delta E_{\rm ring}}{N}
 &=&-\frac{1}{4}\sum_{\alpha\in\{\rm c,T\}}
\int \frac{d^3q}{(2\pi)^3\rho} \int_0^\infty\frac{d\omega}{\pi}
\int_0^1 d\lambda
\Im\frac{\left[\tilde V_{\rm p-h}^{\LSsup}(q)\right]^2\chi_0(q;\omega)\chi_0^{(\perp)}(q,\omega)}
        {1-\lambda \tilde V^{(\alpha)}_{\rm p-h}(q,\omega)\chi_0(q;\omega)}\nonumber\\
        &=&\frac{1}{4}\sum_{\alpha\in\{\rm c,T\}}
        \int \frac{d^3q}{(2\pi)^3\rho} \int_0^\infty\frac{d\omega}{\pi}
        \frac{\left[\tilde V_{\rm p-h}^{\LSsup}(q)\right]^2\chi_0^{(\perp)}(q,\omega)}
             {\tilde V^{(\alpha)}_{\rm p-h}(q,\omega)}
             \ln\left[
               1-\tilde V^{(\alpha)}_{\rm p-h}(q,\omega)\chi_0(q;\omega)\right]\,.
\label{eq:ELSring}
\end{eqnarray}
\end{widetext}

\subsection{Inclusion of exchange diagrams}
\label{ssec:exchange}

\begin{figure}[H]
\centerline{\includegraphics[width=0.7\columnwidth]{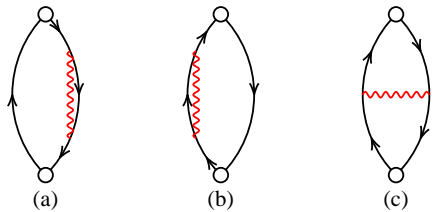}}
\caption{Goldstone diagrams representing the first-order
  corrections to the RPA using the usual conventions of Goldstone
  perturbation theory. Diagram (a) is a self-energy correction to a
  hole propagator, diagram (b) a correction to the particle propagator,
  whereas diagram (c) is the first correction coming from including
  exchanges in the RPA. The wiggly red line represents an
  interaction.
  \label{fig:selfen}}
\end{figure}

Corrections to the RPA are often discussed in terms of two different
effects: One is the modification of the single-particle Green's
functions by self-energy corrections, and the second is the inclusion
of exchange terms of the particle-hole interaction.
Fig. \ref{fig:selfen} shows the lowest order terms in the conventions
of Goldstone perturbation theory. Fig. \ref{fig:selfen}a is the
contribution of the Hartree-Fock exchange term to the hole propagator,
Fig. \ref{fig:selfen}b to the particle propagator, and
Fig. \ref{fig:selfen}c shows an RPA exchange diagram.

Corrections to the single-particle Green's function are, or course,
much more easily treated than exchange terms because one is dealing
with a one-body quantity. The treatment of exchanges is more
complicated, see for example
Ref. \onlinecite{PhysRevC.40.960}. Therefore, the two effects are
normally treated separately, often just leaving one of them out.

\begin{figure}[H]
\centerline{\includegraphics[width=0.85\columnwidth]{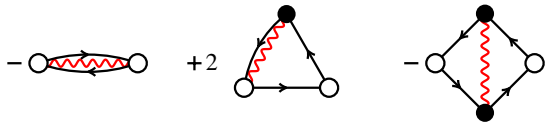}}
\caption{First-order exchange diagrams in the FHNC diagrammatic
  language. Oriented solid lines represent FHNC exchange lines, the
  wiggly red line represents either a correlation or effective
  interaction line \label{fig:eelink}.}
\end{figure}

It is, of course, well known that {\em both\/} contributions must be
retained to satisfy the energy weighted sum rule.  The diagrammatic
analysis of Jastrow-Feenberg wave functions comes to a similar
picture: Fig. \ref{fig:eelink} shows three diagrams of the
Jastrow-Feenberg variational theory for Fermions that have the same
momentum flux as the Goldstone diagrams of Fig. \ref{fig:selfen}. In
fact, the analysis of the diagrams of the Jastrow-Feenberg variational
theory in terms of Goldstone diagrams \cite{RIP79,fullbcs} comes to
the conclusion that the diagrams shown in Fig. \ref{fig:eelink} are
indeed approximations to the Goldstone diagrams shown in
Fig. \ref{fig:selfen}.

From the point of view of Jastrow-Feenberg theory there is ample
evidence that there is a strong cancellation between the three
diagrams shown in Fig. \ref{fig:eelink}. Among others, keeping these
terms together is a necessary condition for obtaining meaningful
solutions of the variational problem \eqref{eq:euler}
\cite{EKVar,polish}.

Concluding that these diagrams should either all be kept or all be
left out, we have dealt with exchange effects as described in
Refs. \onlinecite{v3eos} and \onlinecite{fullbcs}: The first order
exchange diagram depends rigorously on both hole momenta $\hvec$,
$\hvec'$ and the momentum transfer $\qvec$
\begin{equation}
  W_{\rm ex}(\hvec,\hvec';\qvec)
  = \Omega\bra{\hvec+\qvec,\hvec'-\qvec} W \ket{\hvec',\hvec}\,.
  \label{eq:Wex}
\end{equation}

We have approximated this non-local quantity by a function of momentum
transfer by calculating the Fermi-sea average
\begin{widetext}
\begin{equation}
  \tilde V_{\rm ex}(q)\equiv \left\langle  W_{\rm ex}\right\rangle(q)
  = \frac{\sum_{\hvec\hvec'} n(h) n(h') \bar n(|\hvec+\qvec|)
    \bar n(|\hvec'-\qvec|)\Omega\bra{\hvec+\qvec,\hvec'-\qvec}
    W \ket{\hvec',\hvec'}}
       {\left(\sum_{\hvec} \bar n(|\hvec+\qvec|) n(h)\right)^2}\,.\label{eq:favg}
\end{equation}
\end{widetext}
In fact, the averaging procedure \eqref{eq:favg} is used at several
places to establish the connection between the Jastrow-Feenberg
variational theory and the parquet-diagram summations. More details on
the procedure and how the Euler equations are modified by the
inclusion of exchange diagrams are discussed in
Refs. \onlinecite{fullbcs} and \onlinecite{v3eos}.

\section{Results and Summary}
\label{sec:results}
\subsection{Interactions}
We have mostly employed for our calculations the Argonne \cite{AV18}
interaction because it is formulated in terms of the operator
expansion \eqref{eq:vop}. The Reid \cite{Reid68} has been formulated
only in a $v_6$ form \cite{Day81}. To generate a $v_8$ version of the
Reid interaction we have followed the procedure of Ref. \onlinecite{AV18}.
For the sake of discussion, let
$v_\alpha^{(n)}(r_{ij})$ the coefficient of the operator $\hat O_\alpha(i,j)$
in a $v_n$ representation of the interaction. Ref. \onlinecite{AV18}
generates a $v_6$ approximation from a $v_8$ form by
\begin{subequations}
\begin{eqnarray}
  V^{(6)}_{\rm cc}(r) &=& V^{(8)}_{\rm cc}(r) -\frac{0.9}{16}(V_{\rm c,LS}(r) - 3V_{\rm \tau,LS}(r)),
  \label{eq:Vcc}\\
  V^{(6)}_{\rm c\tau}(r) &=& V^{(8)}_{\rm c\tau}(r) +
  \frac{0.9}{16}(V_{\rm c,LS}(r) - 3V_{\rm \tau,LS}(r)),\label{eq:Vct}\\
  V^{(6)}_{\rm \sigma c}(r) &=& V^{(8)}_{\rm \sigma c}(r) -
  \frac{0.3}{16}(V_{\rm c,LS}(r) - 3V_{\rm \tau,LS}(r)),\label{eq:Vsc}\\
  V^{(6)}_{\rm \sigma\tau}(r) &=& V^{(8)}_{\rm \sigma\tau}(r) +\frac{0.3}{16}(V_{\rm c,LS}(r)
  - 3V_{\rm \tau,LS}(r)).\label{eq:Vst}
\end{eqnarray}
\end{subequations}
We have used this procedure to generate a $v_8$ representation of the
Reid interaction as follows: The spin-orbit potential components
$V_{c,LS}(r)$ and $V_{\tau,LS}(r)$ were constructed from the isospin=0
and isospin=1 components of the Reid interaction, \cf Eqs. (20) and
(30) of Ref. \onlinecite{Reid68}. The four components
$V^{(6)}_{\alpha}(r)$, $\alpha \in \{({\rm cc}), (\rm{c}\tau), (\sigma{\rm c}),
(\sigma\tau)\}$ were taken from Eqs (A3)-(A8) of
Ref. \onlinecite{Day81}. Then we used Eqs. \eqref{eq:Vcc}-\eqref{eq:Vst}
to construct a $v_8$ representation of the Reid interaction.

Another interaction that is given in the operator basis is the one of
Wiringa, Smith and Ainsworth \cite{Wiri84}. This interaction has been
used by Smith and Jackson \cite{SmithSpin} in a $v_6$ calculation of a
fictitious system of interacting bosons.  These authors argue that the
spin-orbit interaction can indeed have a visible effect in nuclear
matter.  We have not been able to verify this in neutron matter and
leave the issue to further study.

\subsection{Long wavelength limit}
\label{ssec:qzero}
The most important input for linear response theory and, hence, for
the calculation of the dynamic structure function, is the
particle-hole interaction. The hydrodynamic speed of
sound
\begin{equation}
mc^2 = \frac{d}{d\rho}\rho^2 \frac{d}{d\rho}\frac{E}{N}\,.
\label{eq:mcfromeos}
\end{equation}
is related to the long-wavelength limit of the particle-hole
interaction. In a Fermi fluid, we also have Pauli repulsion, reflected
in the relation
\begin{equation}
mc^2 = mc_{\rm F}^{*2} + \tilde V_{\rm p-h}(0+) \equiv  mc_{\rm F}^{*2}(1+F_0^S)
\,,
\label{eq:FermimcfromVph}
\end{equation}
where $c_{\rm F}^* = \sqrt{\frac{\hbar^2\KF^2}{3mm^*}}$ is the speed
of sound of the non-interacting Fermi gas with the effective mass
$m^*$, and $F_0^s$ is Landau's Fermi liquid parameter. In neutron
matter we can safely set $m^*=m$ \cite{ectpaper}. We note in passing
that the local approximation \eqref{eq:favg} gives the correct
contribution to $mc^2$ as defined in Eq. \eqref{eq:FermimcfromVph}
from the leading-order exchange diagrams \cite{EKVar}.

The relationships (\ref{eq:mcfromeos}) and (\ref{eq:FermimcfromVph})
give identical predictions only in an exact theory
\cite{EKVar,parquet5}; good agreement is typically reached only at
very low densities. Even in the much simpler system $^4$He, where
four- and five-body elementary diagrams and three-body correlations
are routinely included, the two expressions (\ref{eq:mcfromeos}) and
(\ref{eq:FermimcfromVph}) can differ by up to a factor of two
\cite{lowdens}.

The situation is even more complicated in Fermi systems again due to
the multitude of exchange diagrams, of which we kept only the
simplest.  Hence, one can expect good agreement only at very low
densities \cite{fullbcs}, but not at the densities considered here.
To make valid predictions on the density-density response function we
have followed here the procedure of Ref. \onlinecite{lowdens} and
added a phenomenological component to the theory: We have scaled, for
$q\le 2\KF$ the exchange correction $\tilde V_{\rm ex}(q)$ such that
the two expressions (\ref{eq:mcfromeos}) and (\ref{eq:FermimcfromVph})
agree.  The procedure has no visible effect on the equation of state.
One can think of a similar procedure for $\tilde V_{\rm
  p-h}^{\rm (L)}(0+)$, by comparing the microscopically calculated Landau
  parameter with a magnetic susceptibility obtained by calculating the
  equation of state for partially spin-polarized systems.

\subsection{Effective spin-orbit interaction}
\label{ssec:WLS}

One of the main focuses of interest is, of course, the consequences of
many--body correlations on effective interactions. We have introduced
above the ``particle-hole interaction'' $\hat V_{\rm p-h}(\qvec)$ (\cf
Eq.  \eqref{eq:VphLS}), the effective interactions $\hat
W(\qvec;\omega)$, and the induced interaction $\hat W_{\rm
  I}(\qvec;\omega)$, Eq. \eqref{eq:WindRPA}, as well as their
energy-independent counterparts.

The  tensor  force  breaks,  in   the  spin-triplet  case  $S=1$,  the
degeneracy  of  the  correlations  in  the  $M_S=1$  and  the  $M_S=0$
channels,  described  by  $\Gamma^{(t+)}(r)$  and  $\Gamma^{(t-)}(r)$.
This  would lead  to  {\em two  different\/} spin-orbit  particle-hole
interactions  for $M_S=0$  and $M_S=1$,  see Eq.  \eqref{eq:VphLS}. We
have taken  in Eq.  \eqref{eq:VphLS} generally the  $M_S=1$ projection
for two reasons: For most of the calculations to follow, the influence
of  the  spin-orbit   interaction  is  small.  This   is  because  the
corrections  to  the  effective  interactions  are  quadratic  in  the
spin-orbit    potential,     see    Eqs.     \eqref{eq:Vcredef}    and
\eqref{eq:VTredef}. We found that using  the $M_S=0$ projection of the
correlation leads  to an even  smaller correction from  the spin-orbit
terms. The second reason is  that the likely most prominent influence
of the  spin-orbit effective  interaction is  on $^3\!P_2$  pairing in
high-density neutron matter. In that  case, the $M_S=1$ channel is the
appropriate one. We also stress that, in this case, neither of the
local  approximations  discussed  in Sec.  \ref{sssec:LocalChains}  is
necessary; in fact  it is more appropriate to  calculate the effective
interaction for $\hbar\omega=0$ \cite{v3bcs}.

\begin{figure}[H]
  \centerline{\includegraphics[width=0.55\columnwidth,angle=-90]%
      {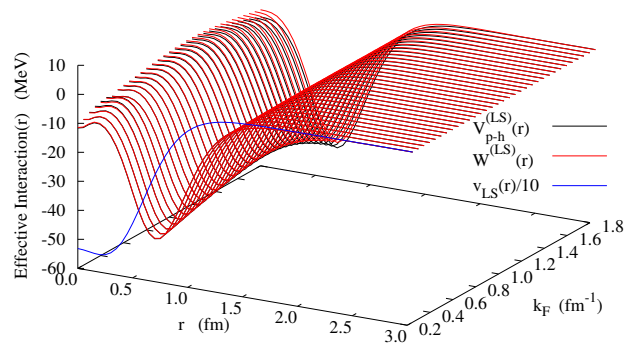}}
  \caption{The figure shows the particle-hole interaction
    $V^{\LSsup}_{\rm p-h}(r)$ (black lines), the effective interaction
    $W^{\LSsup}(r)$ (red lines) and, for comparison, the bare
    spin-orbit interaction (blue line). The bare interaction has been
    scaled by a factor 0.1 to fit into the plot.\label{fig:VLSplotarg}}
\end{figure}

Fig. \ref{fig:VLSplotarg} shows the particle-hole interaction $V_{\rm
  p-h}^{\LSsup}(r)$, the static effective interaction $W^{\LSsup}(r)$,
and the bare spin-obit potential \cite{AV18}. Evidently, the full
effective interaction and the particle-hole interaction are almost
indistinguishable, their difference is certainly less than the
accuracy of the parquet//1 approximation used here. That says that
``induced interaction'' effects to the spin-orbit component of the
effective interaction are negligible. This does of course not imply
that the induced interaction is negligible for the other components of
the interaction, see Figs. 10 and 11 of Ref. \onlinecite{v3eos}.

The most prominent effect that modifies the bare interaction in the
spin-orbit channel is the screening of its short-ranged behavior by
the correlations caused by the surrounding particles. This screening
is manifested in the factor $1+\Gamma^{(t+)}(r)$ in
Eq. \eqref{eq:VphLS}. Thus, our result is that the effective
interaction has very little resemblance to the bare interaction, but
that a relatively simple treatment of correlations is adequate to deal
with many-body effects.

\begin{figure}[H]
  \centerline{\includegraphics[width=0.65\columnwidth,angle=-90]%
      {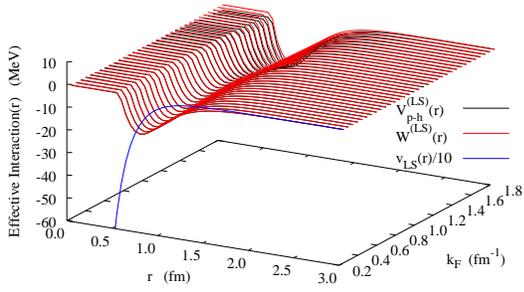}}
  \caption{Same as Fig. \ref{fig:VLSplotarg} for our $v_8$ formulation
    of the Reid interaction.\label{fig:VLSplotrei}}
\end{figure}

The situation is very similar for the $v_8$ representation of the Reid
potential, see Fig. \ref{fig:VLSplotrei}.  The spin-orbit component of
the Reid interaction is singular as $r\rightarrow 0$, thus it appears
to be much stronger than the bare spin-orbit interaction from the
Argonne potential. However, the short-ranged screening is also
stronger, resulting in an effective interaction that is similar to
the one derived from the Argonne potential.

\begin{figure}[H]
  \centerline{\includegraphics[width=0.55\columnwidth,angle=-90]{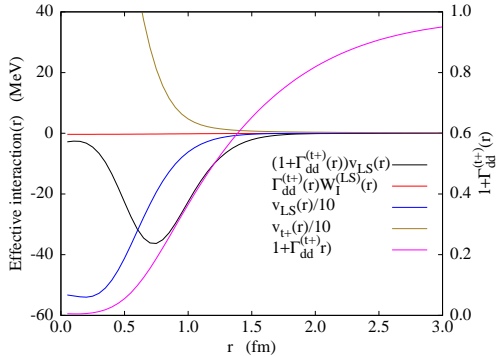}}
  \caption{The figure shows, for $\KF = 1.0\,\text{fm}^{-1}$
    the individual contributions to the effective spin-orbit
    interaction (left scale). We also show the direct correlation
    function $1+\Gamma_{\rm dd}^{(t+)}(r)$ in the $t+$ projector
    channel (magenta line, right scale) and the bare interactions
    $v_{\rm LS}(r)$ and $v_{t+}(r)$ of the Argonne potential. These
    were scaled by a factor of 0.1 to fit into the plot.
    \label{fig:vkf100arg}}
\end{figure}

\begin{figure}[H]
  \centerline{\includegraphics[width=0.55\columnwidth,angle=-90]{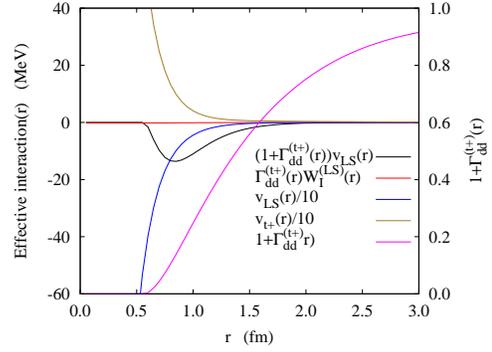}}
  \caption{Same as Fig. \ref{fig:vkf100arg} for the Reid interaction.
    \label{fig:vkf100rei}}
\end{figure}

More details on the different contributions to the effective
interactions are shown in Fig. \ref{fig:vkf100arg} and
\ref{fig:vkf100rei}.  The figures show, at the representative density
$\KF = 1.0\,\text{fm}^{-1}$ the decomposition \eqref{eq:VphLS} of the
particle-hole interaction into the short-ranged screening effect
$v_{\rm LS}(r)(1+\Gamma^{(t+)}_{\rm dd}(r))$ and the induced
interaction $\Gamma^{(t+)}_{\rm dd}(r)W_I^{\LSsup}(r)$. We also show,
for comparison, the bare interactions $v_{\rm LS}(r)$ and $v_{t+}(r)$.
The comparison offers an explanation for why the spin-orbit potential
is strongly suppressed by many-body correlations. The pair correlation
$1+\Gamma^{(t+)}_{\rm dd}(r)$ is predominantly determined by the
$v_{t+}(r)$ which is strongly repulsive. Hence, the correlation
function tends to suppress the interaction.

Fig. \ref{fig:vkf100rei} shows the same data for the $v_8$ version of
the Reid potential. On the other hand, the $t+$ component
of the Rein interaction is slightly more repulsive than that of the
Argonne potential, causing a somewhat stronger short-ranged screening.
As a result, the spin-orbit component of the effective interactions
is even smaller than that of the Argonne potential.

\subsection{Corrections to the particle-hole interactions in the central
  and transverse channel}

\begin{figure}[H]
  \centerline{\includegraphics[width=0.6\columnwidth,angle=-90]{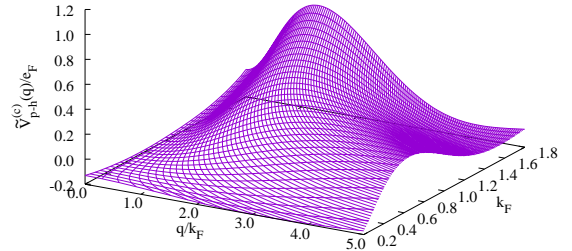}}
  \caption{The figure shows the central component
    $\tilde V^{\rm (c)}_{\rm p-h}(q)$ of the particle-hole interaction
    in units of the Fermi energy of the non-interacting system.
\label{fig:Vph1}}
\end{figure}

\begin{figure}[H]
  \centerline{\includegraphics[width=0.6\columnwidth,angle=-90]{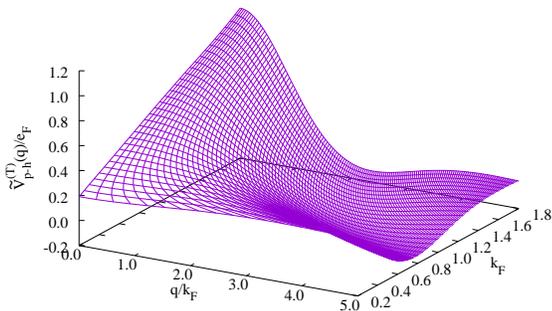}}
  \caption{Same as Fig. \ref{fig:Vph1} for the transverse component
    $\tilde V^{\rm (T)}_{\rm p-h}(q)$ of the particle-hole interaction.\label{fig:VphT}}
\end{figure}

The essential input to the calculation of the density-density response
function is the particle-hole interaction. The spin-orbit potential
contributes a relatively small energy-dependent correction. Figs.
\ref{fig:Vph1} and \ref{fig:VphT} show the $\tilde V^{\rm (c)}_{\rm
  p-h}(q)$ and $\tilde V^{\rm (T)}_{\rm p-h}(q)$, the results differ visibly from
those obtained in Ref. \onlinecite{v3eos} by the inclusion of
non-parquet ``twisted chain'' contributions and the modification
of the exchange interaction discussed in section \ref{ssec:qzero}.


\begin{figure}[H]
  \centerline{\includegraphics[width=0.6\columnwidth,angle=-90]{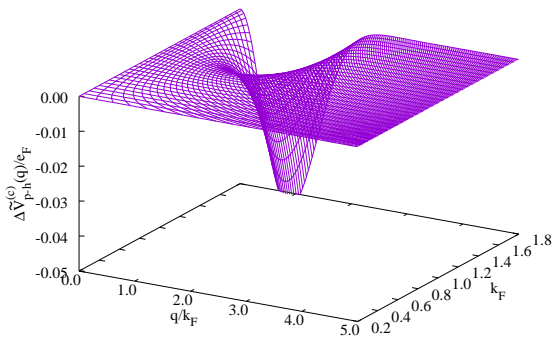}}
  \caption{The figure shows  the correction
    spelled out in Eq. \eqref{eq:Vcredef} to the central part of the 
    particle-hole interaction \ref{fig:Vph1} taken at the average energy
    $\hbar\omega_F(q) = t(q)/\SF(q)$. \label{fig:VphLS}}
\end{figure}

Fig. \ref{fig:VphLS} finally shows the correction \eqref{eq:Vcredef} due
to spin-orbit interaction, note that the correction to the transverse
channel is half of that, see Eq. \eqref{eq:VTredef}. Evidently the
correction is very small, this is due to two effects, namely that
the spin-orbit correction is of second order, and that the spin-orbit
force gets effectively screened by many-body correlations of the
surrounding medium, see Fig. \ref{fig:vkf100arg}.

We conclude this section by remarking that the correction \eqref{eq:ELSring}
has turned out to be less than 0.1 MeV which is of the order of a percent
of the total energy. It is therefore not shown here.

\subsection{Dynamic structure function}
\label{ssec:skw}

In what follows we focus on results obtained for the Argonne interaction,
those for the Reid potential are quire similar and nothing can be learned
from a comparison.

The dynamic structure function of nuclear and neutron matter has been
the subject of intense studies literally for decades, see among others
Ref. \onlinecite{ALBERICO1982429} for very early work. These
calculations were typically at the RPA level, interactions were taken
either semi-phenomenologically using simplified nucleon-nucleon
interactions \cite{ALBERICO1982429,Hensel1983,PhysRevC.40.960},
effective Skyrme interactions
\cite{PhysRevC.80.024314,PhysRevC.84.059904,PhysRevC.100.064301,%
  PhysRevC.89.044302,JPG41_2014,PhysRep536,AnnPhys214} or
pseudopotentials \cite{Pines:1988gik}. Closest to our formulation and
philosophy is the pseudopotential method. In fact, the
phenomenological considerations that went into the construction of
pseudopotentials for the helium liquids \cite{Aldrich,ALP78} are
faithfully reproduced and thereby justified by microscopic
calculations \cite{EKthree,polish}. Developing pseudopotentials is of
course much more complicated in neutron and nuclear matter due to the
absence of extensive experimental data; phenomenological theories
therefore suffer from more ambiguities. There are only a few low-order
calculations that attempt to include correlation effects in the
dynamic response \cite{NaiThesis,Benhar2009,Lovato2013}.

In our microscopic calculations, the response of the system to the
external fields \eqref{eq:extfields} is naturally formulated in terms
of the operators $\1$, $\hat L$ and $\hat T$, see
Eq. \eqref{eq:chialpha}. The tensor force breaks the degeneracy in
$\hat L$ and $\hat T$.

\begin{figure}[H]
  \centerline{\includegraphics[width=0.5\columnwidth,angle=-90]{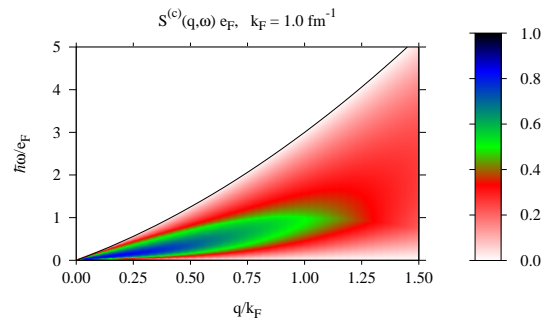}}
  \caption{(color online) The figure shows a color map of the density
    channel $S^{\rm (c)}(q,\omega)$ of the dynamic structure function at
    $\KF = 1.0\,\text{fm}^{-1}$. The solid line is the upper boundary
    of the particle-hole continuum. \label{fig:SKW3DC}}
\end{figure}

\begin{figure}[H]
  \centerline{\includegraphics[width=0.5\columnwidth,angle=-90]{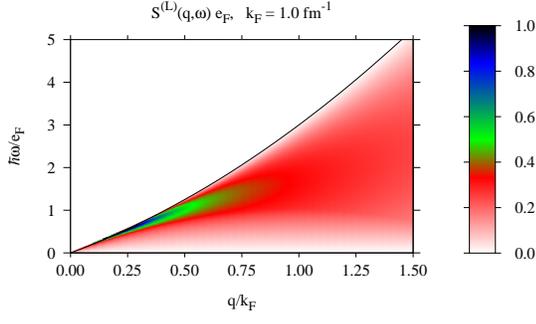}}
  \caption{(color online) Same as Fig. \ref{fig:SKW3DC} for the
    longitudinal channel $S^{\rm (L)}(q,\omega)$ of the dynamic structure
    function. \label{fig:SKW3DL}}
\end{figure}

\begin{figure}[H]
  \centerline{\includegraphics[width=0.5\columnwidth,angle=-90]{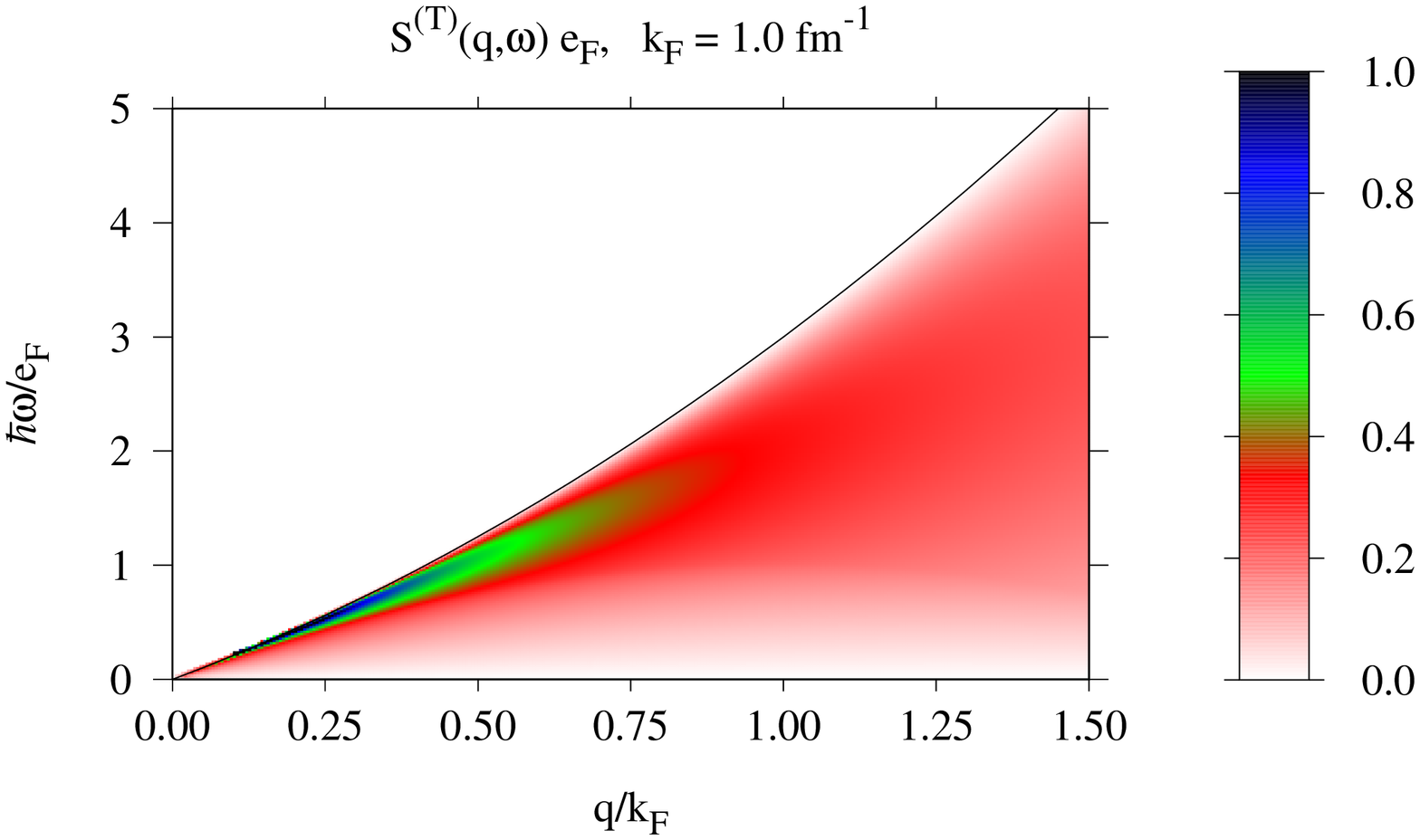}}
  \caption{(color online) Same as Fig. \ref{fig:SKW3DC} for the
    transverse channel $S^{\rm (T)}(q,\omega)$ of the dynamic structure
    function. \label{fig:SKW3DT}}
\end{figure}

A first overview of our results is given in Figs. \ref{fig:SKW3DC},
\ref{fig:SKW3DL} and \ref{fig:SKW3DT}.  We show there results for $\KF
= 1.0\,\text{fm}^{-1}$, the results for different densities are rather
similar. The most evident difference between the distinct channels is
that the strength of $S^{\rm (c)}(q,\omega)$ is mostly in the middle of
the particle-hole continuum whereas $S^{\rm (L)}(q,\omega)$ and
$S^{\rm (T)}(q,\omega)$ show, at long wavelengths, significant strength
just below the boundary of the particle-hole continuum. Our
results are in that aspect similar to those obtained with Skyrme
interactions \cite{PhysRep536,AnnPhys214} and at low cluster
orders \cite{Lovato2013} who occasionally find that mode outside the
continuum.


\begin{figure}[H]
  \centerline{\includegraphics[width=0.70\columnwidth,angle=-90]{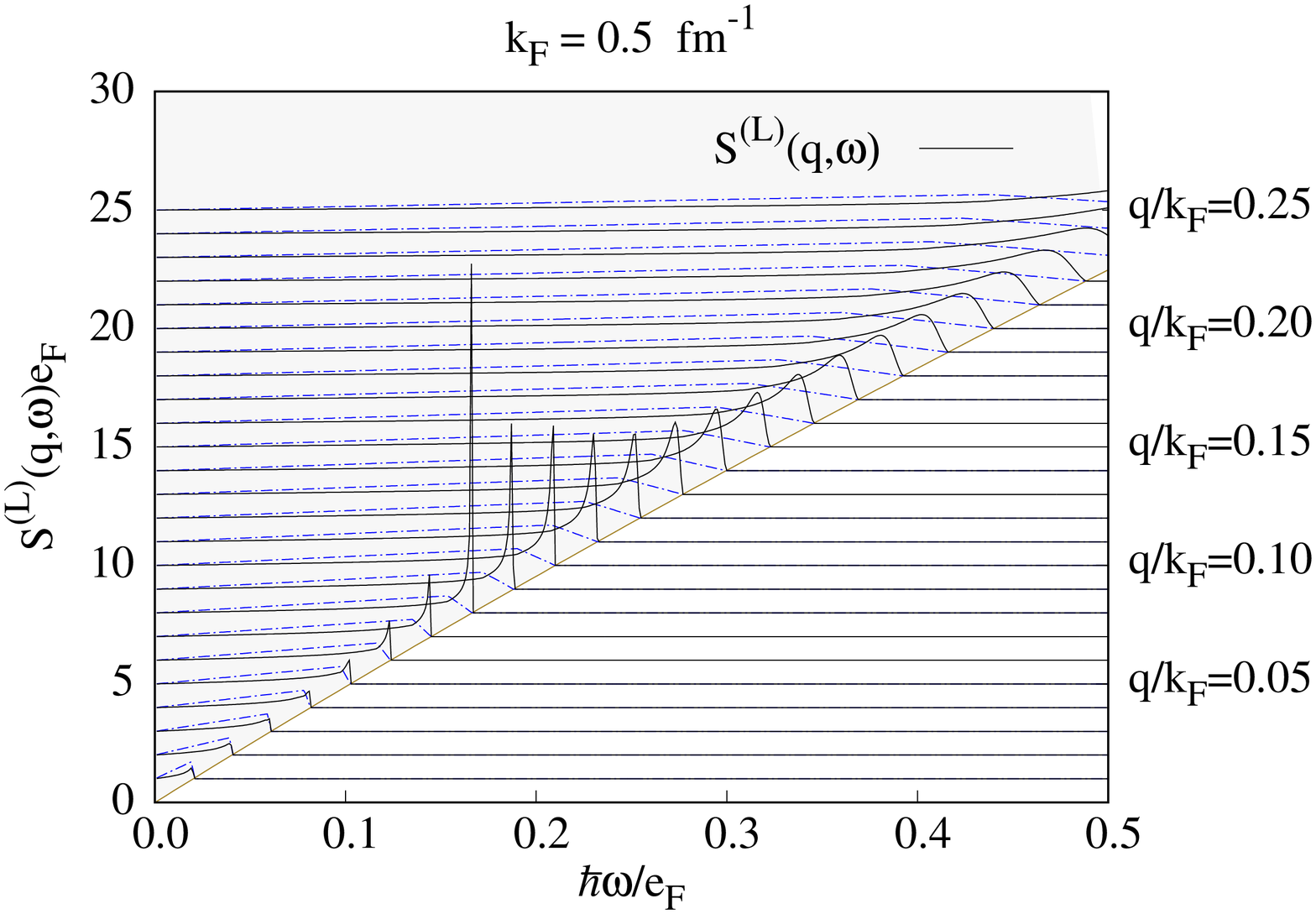}}
  \centerline{\includegraphics[width=0.70\columnwidth,angle=-90]{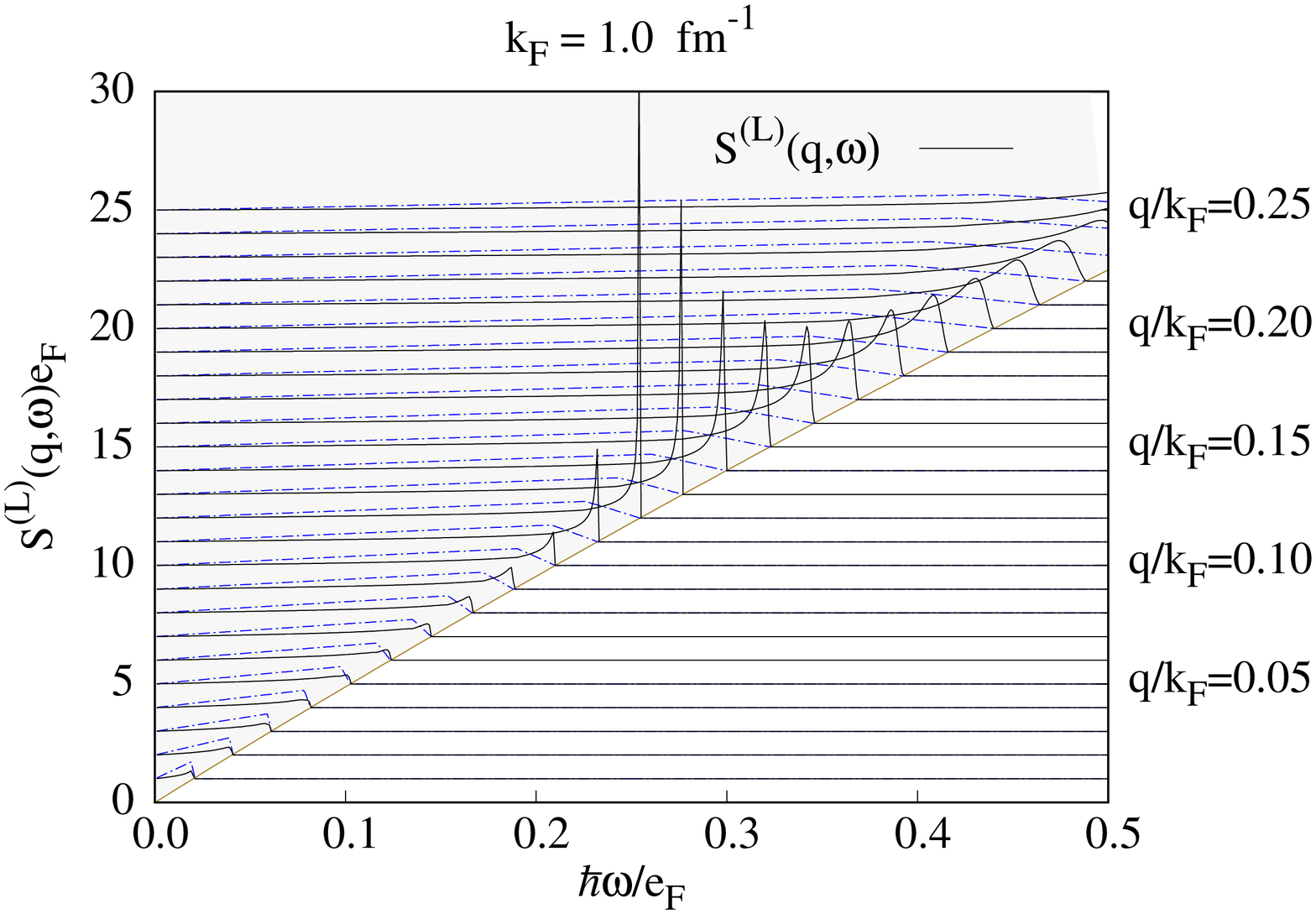}}
  \centerline{\includegraphics[width=0.70\columnwidth,angle=-90]{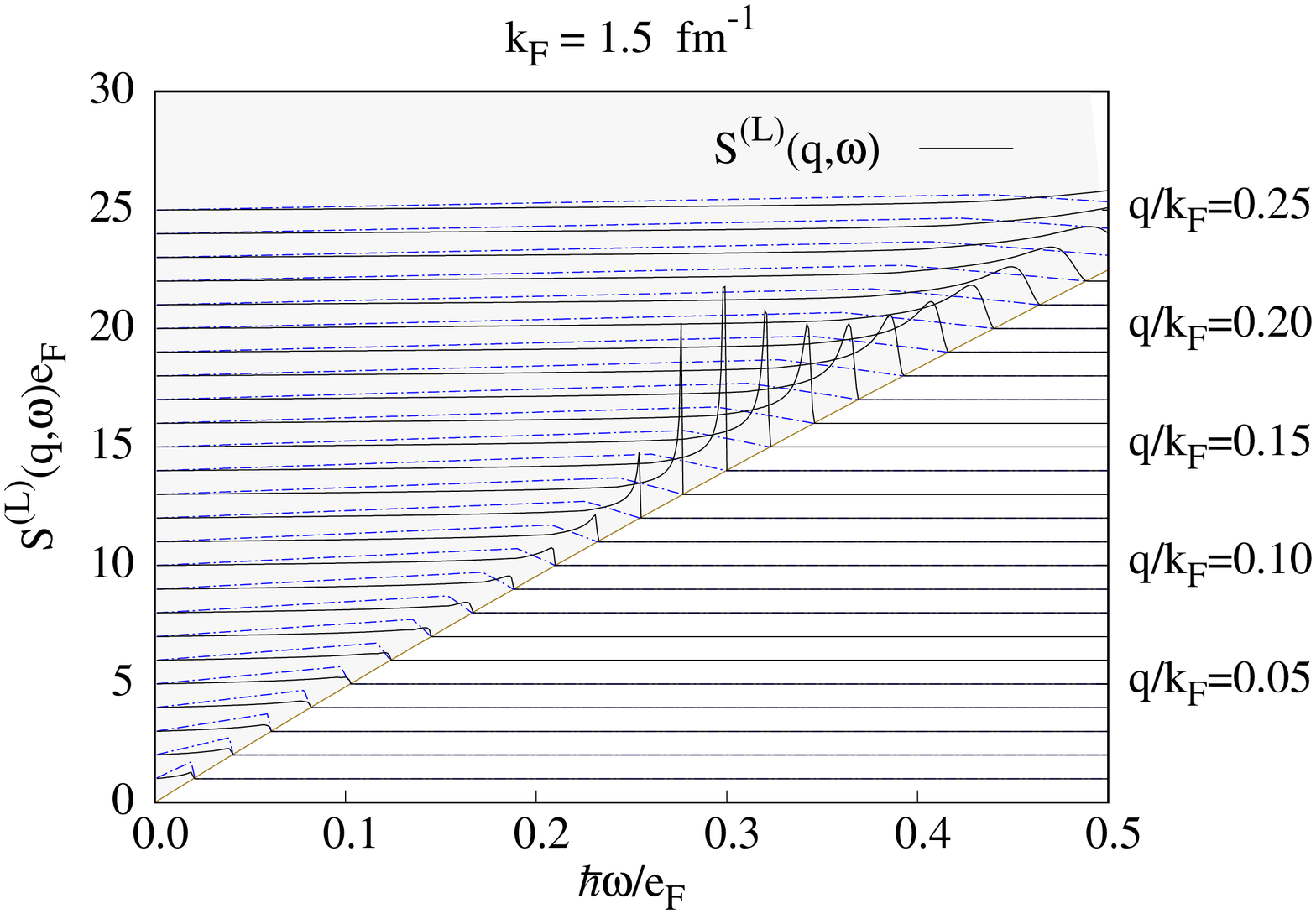}}
  \caption{(color online) The figures show details of
    $S^{\rm (L)}(q,\omega)$ for three densities as indicated in the plots for momentum
    transfers $q/\KF = 0.01\,\ldots\,0.25$. The individual results are
    stacked as indicated on the right scale of the plots. The dashed
    blue lines show, for comparison, the dynamic structure function of
    the non-interacting system, and the gray-shaded area shows the
    particle-hole continuum.
  \label{fig:SKWLAV18}}
\end{figure}

\begin{figure}[H]
  \centerline{
  \includegraphics[width=0.70\columnwidth,angle=-90]{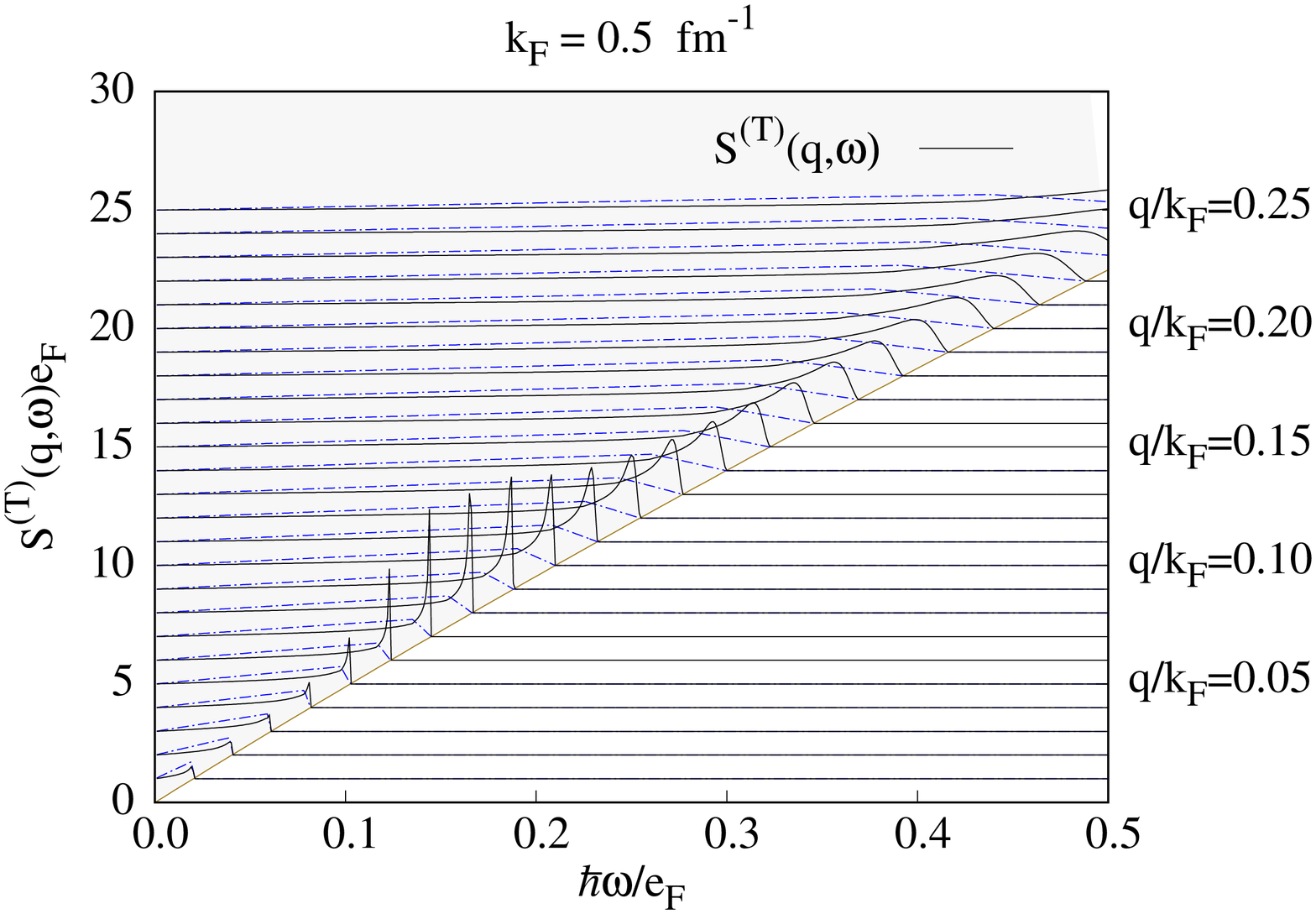}}
  \centerline{
  \includegraphics[width=0.70\columnwidth,angle=-90]{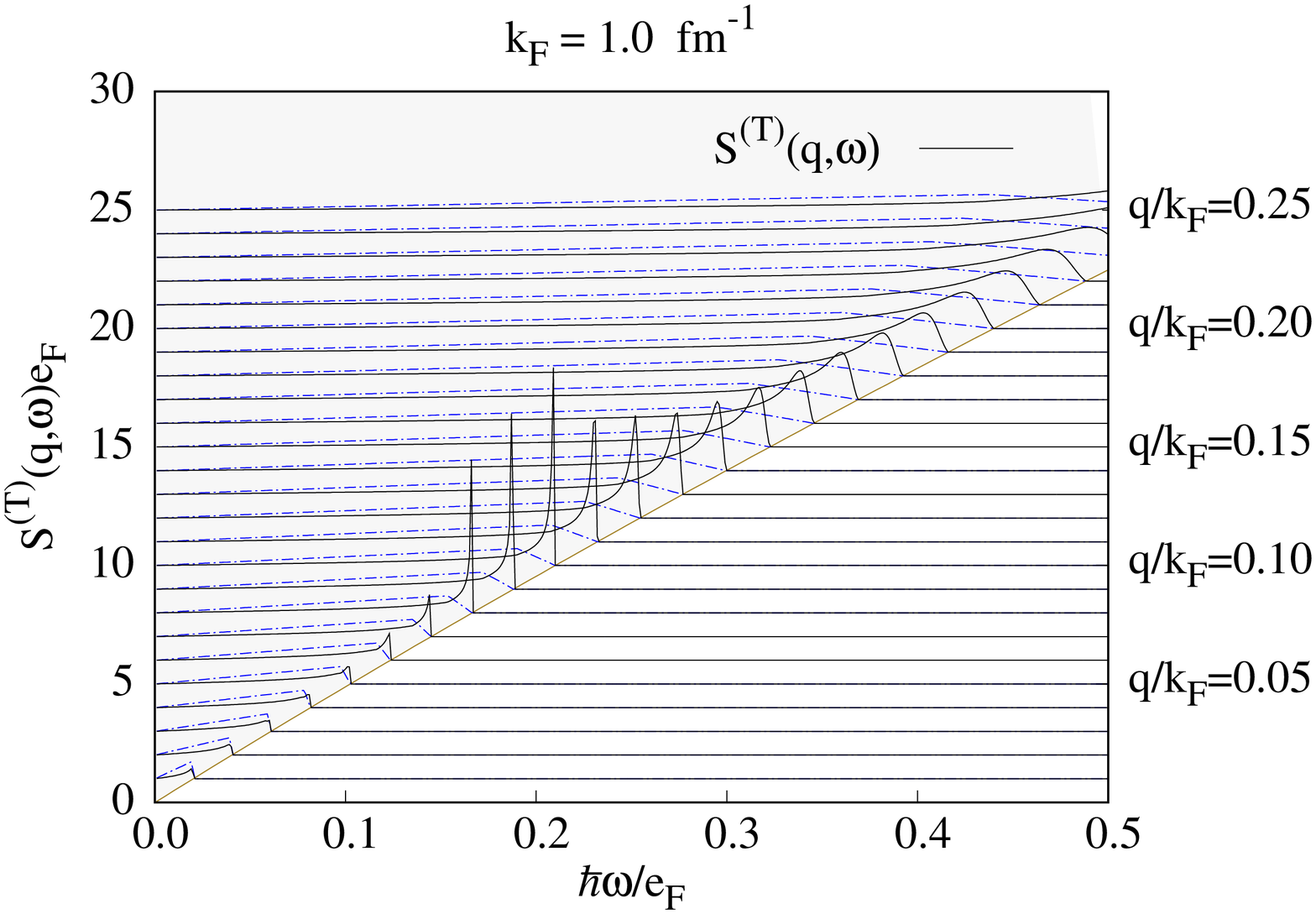}}
  \centerline{
  \includegraphics[width=0.70\columnwidth,angle=-90]{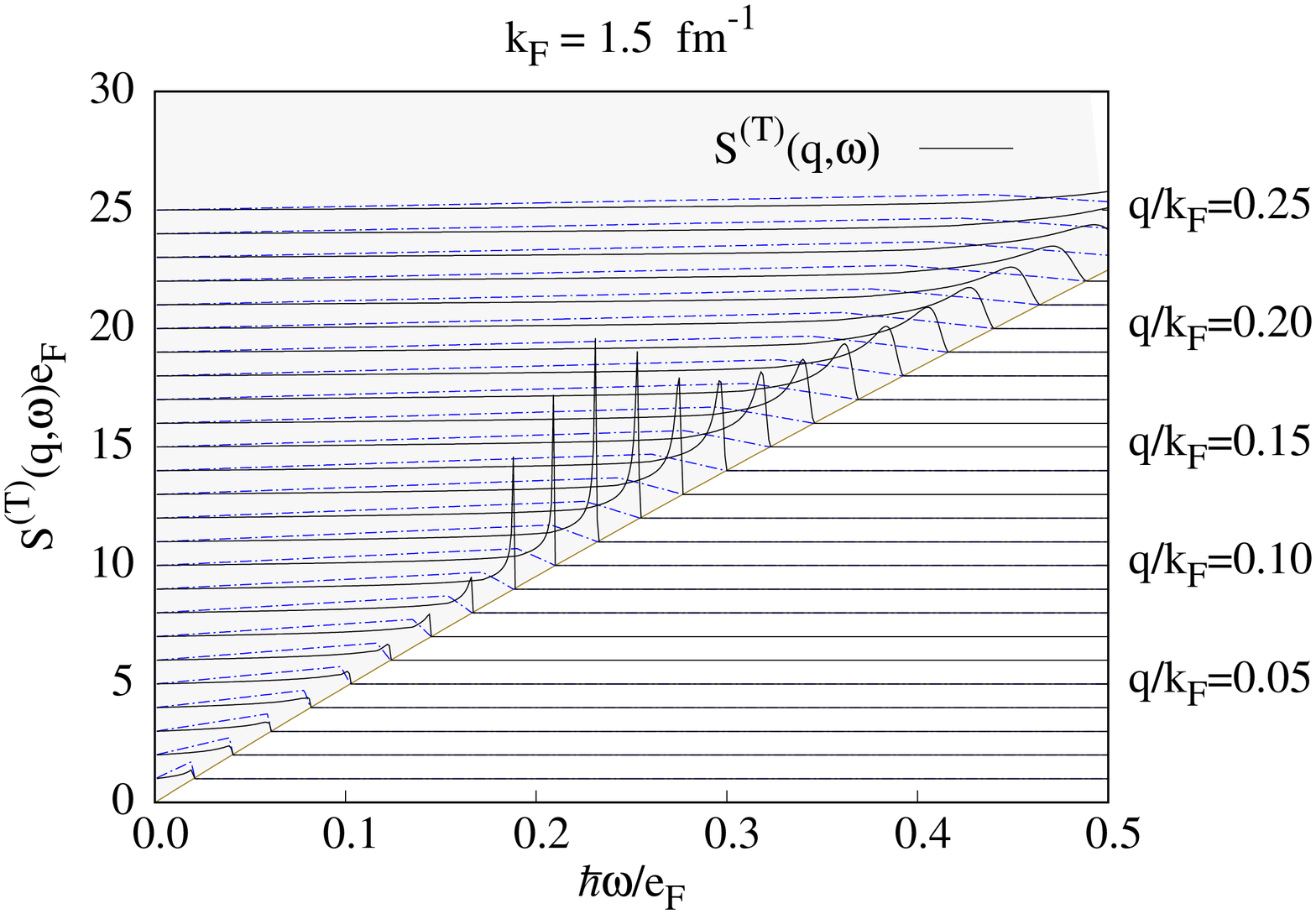}}
  \caption{(color online) Same as Fig. \ref{fig:SKWLAV18} for $S^{\rm (T)}(q,\omega)$. 
  \label{fig:SKWTAV18}}
\end{figure}

Let us therefore have a closer look at the long wavelengths properties
at different densities. Details are shown in Figs.
\ref{fig:SKWLAV18} and \ref{fig:SKWTAV18}.
The peaks in both  $S^{\rm (L)}(q,\omega)$ and $S^{\rm (T)}(q,\omega)$ are clearly seen,
the figures also substantiate the remark made above that the results
are rather similar at different densities.


The peak in $S^{\rm (L)}(q,\omega)$ and $S^{\rm (T)}(q,\omega)$ is caused by a
node of the real part of the denominators in Eqs. \eqref{eq:chialpha},
\ie by the solution of

\begin{equation}
  \Re \left[1-\tilde V_{\rm p-h}^{(\alpha)}(q,\omega_{\rm zs}(q))
    \chi_0(q,\omega_{\rm zs}(q))
    \right] = 0\,,\label{eq:ezsdef}
\end{equation}
with $\omega_{\rm zs}(q)$ being the zero-sound mode.

\begin{figure}[H]
  \centerline{\includegraphics[width=0.6\columnwidth,angle=-90]{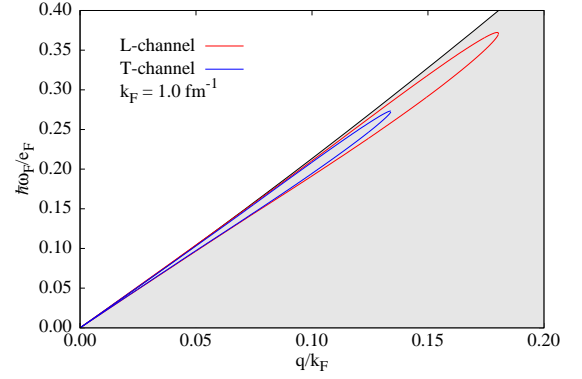}}
  \caption{(color online) The figure shows the location of the
    solution of Eq. \eqref{eq:ezsdef} for $\alpha={\rm L}$ and
    $\alpha={\rm T}$. The gray shaded area is the particle-hole
    continuum. \label{fig:ezsTkf100}}
\end{figure}

In our case, the solutions of Eq. \eqref{eq:ezsdef} are inside the
continuum, that means the imaginary part is non-zero but evidently
very small. Fig. \ref{fig:ezsTkf100} shows, for $\KF =
1.0\,\text{fm}^{-1}$, the location of the ``zero sound pole'' for
longitudinal and transverse excitations.

\begin{figure}[h]
  \centerline{\includegraphics[width=0.6\columnwidth,angle=-90]{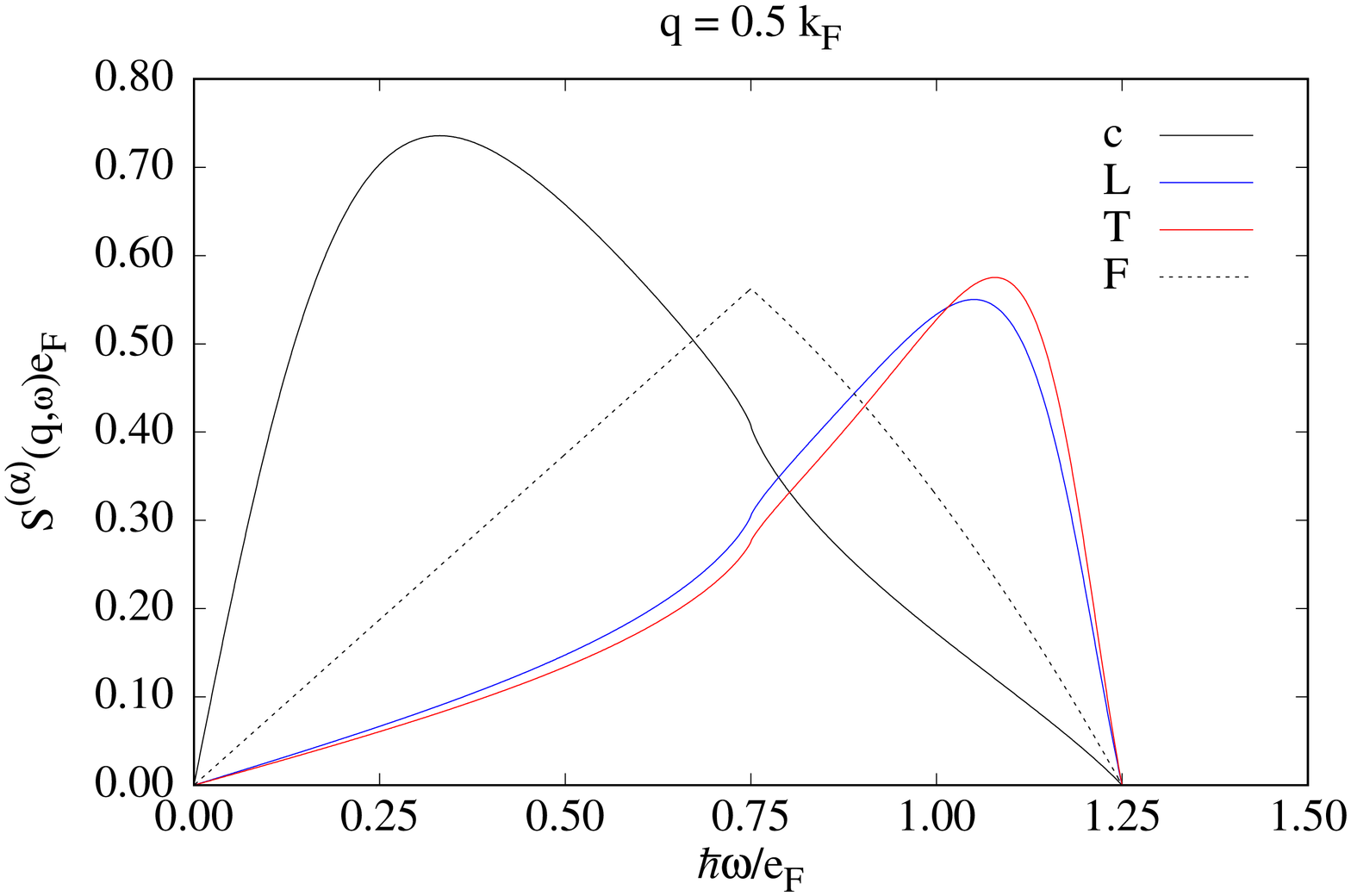}}
  \centerline{\includegraphics[width=0.6\columnwidth,angle=-90]{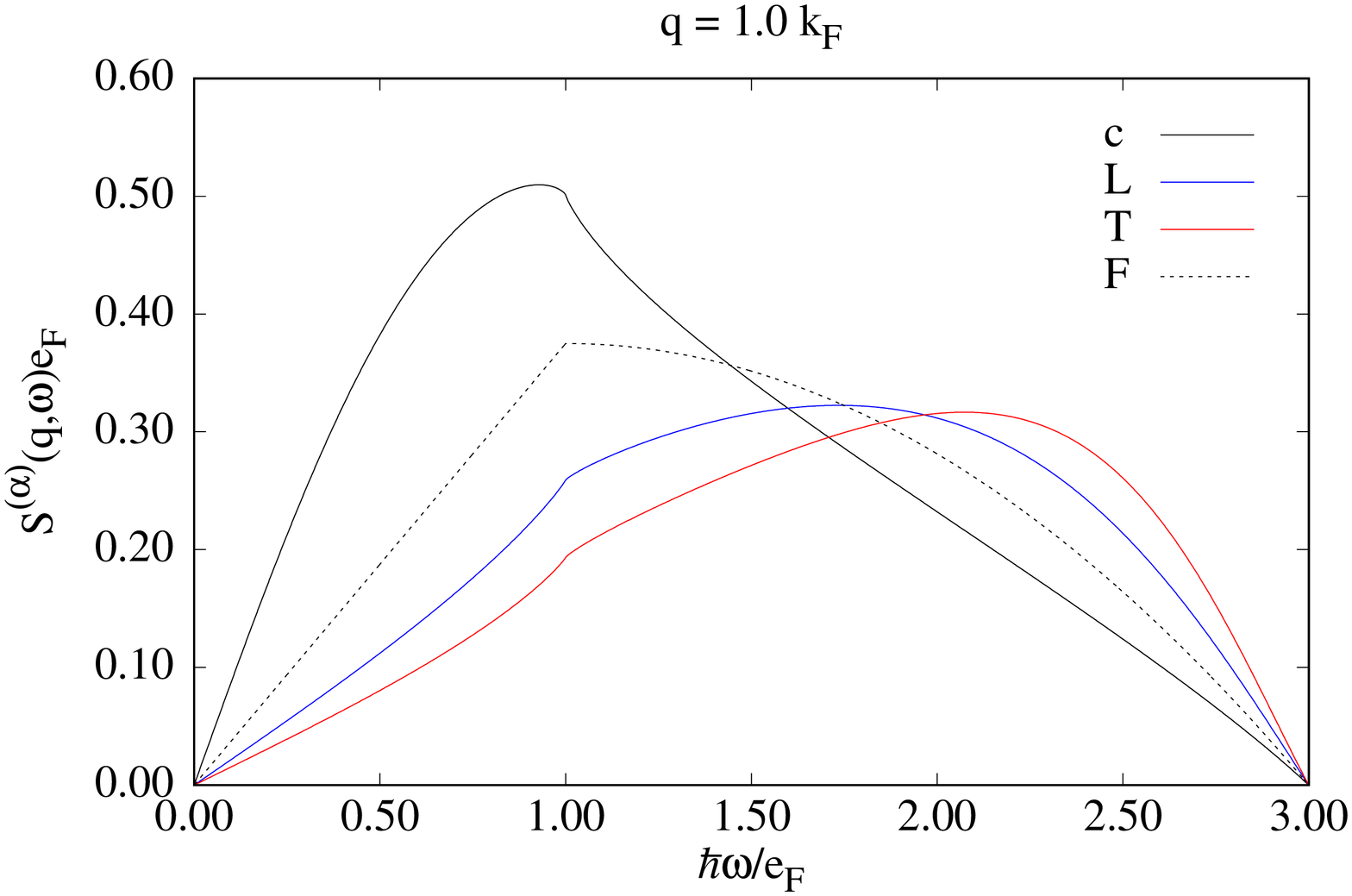}}
  \centerline{\includegraphics[width=0.6\columnwidth,angle=-90]{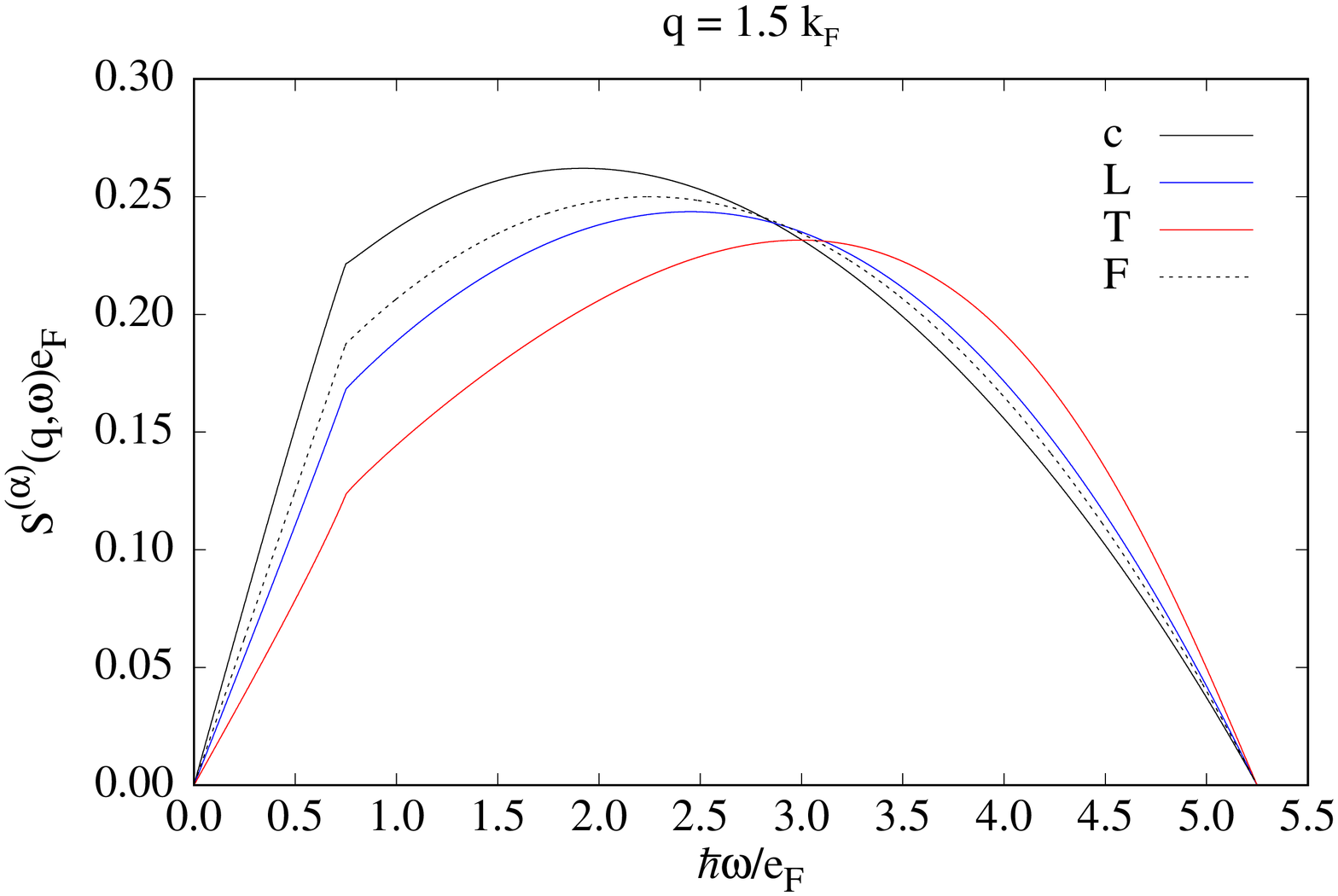}}
  \caption{(color online) The figure shows, for $\KF = 1.0\,\text{fm}^{-1}$,
    the c (black line), L (blue line), and T (red line) components
    of the dynamic structure function at $q = 0.5, 1.0, 1.5\,\KF$.
    The black dashed line is the dynamic structure function of the
    non-interacting Fermi system.\label{fig:SKWcuts}}
\end{figure}

A closer look at the strength distribution in the
$S^{(\alpha)}(q,\omega)$ is provided in
Figs. \ref{fig:SKWcuts}. Similar to what we have seen above, the
strength in the central channel is broadly distributed within the
particle-home continuum whereas the strength of the two spin modes is
shifted towards the upper boundary. This holds quite well up to
$q\approx 0.5\,\KF$ although the node of Eq. \eqref{eq:ezsdef}
disappears at $q < 0.2\,\KF$.  With increasing momentum transfer the
structure functions become closer to the non-interacting limit. The
same holds at other densities.

\newpage
\subsection{Summary}

We have in this work first developed the parquet-diagram summation
method for the $v_8$ model of the nucleon-nucleon
interaction. Our results have been threefold:

We have first derived closed-form expressions for the effective
interaction. These are non-local and are {\em not\/} of the $v_8$ form
but contain three more operators, $\hat Q_7$, $\hat Q_9$ and $\LSp$,
(Eqs. \eqref{eq:Q79def} and \eqref{eq:LSpdef}.). Additionally, these
interactions are energy-dependent.  As long as one just needs these
interactions as, for example, for the examination of pairing phenomena,
there is no need for further approximations. Only for the purpose of
parquet-diagram summations, local approximations to these effective
interactions have been introduced in section \ref{sssec:LocalChains}.

Next we have examined, in section \ref{ssec:WLS}, the effect of
correlations on the spin-orbit interaction. We have demonstrated in
Figs. \ref{fig:VLSplotarg},  \ref{fig:VLSplotrei}  and \ref{fig:vkf100arg}
 \ref{fig:vkf100rei} that many-body
correlations have a rather drastic screening effect. The induced
interaction, which plays a visible role in $S$-wave pairing effects, is
almost negligible in the spin-orbit channel whereas it is quite
visible in the central, longitudinal and transverse channels \cite{v3eos,v3twist}.

Finally, we have joined the long row of works dealing with the
(spin-)density response functions of neutron matter. Our work is
distinguished from most of the previous work in the sense that it is
manifestly microscopic whereas Skyrme interactions or pseudopotentials
require significant phenomenological input which is sometimes hard to
justify in a fictitious system like neutron matter. Closest to out
work are early calculations by Kwong \cite{NaiThesis} and calculations
at the three-body cluster level \cite{Lovato2013}.

We found that, apart from the direct calculation of spin-orbit
effective interaction, the effect of the spin-orbit potential is
rather small. In particular, we found that the energy correction
\eqref{eq:ELSring} is negligibly small.
  \appendix

\begin{widetext}
\section{Details of the chaining operations}
\label{app:vchain}

We present in this appendix details of the calculation of the
effective potential $\hat W(\qvec)$ containing a spin-orbit interaction.
Let us begin with the case that the particle-hole interaction consists
of a spin-orbit term alone; we will then combine our results
with the other channels. The particle-hole
spin-orbit interaction in momentum space is then simply
\begin{equation}
  \hat V_{\rm p-h}^{\LSsup}(\qvec)=\tilde V_{\rm p-h}^{\LSsup}(q)\LS\,.
\end{equation}

\subsection{Chains of pure spin-orbit operators}
  \label{app:vchain2}
We begin with the second order convolution product
\begin{eqnarray}
  \hat W_{\rm LS}^{(2)}(\qvec,\hvec,\hvec',\bsigma,\bsigma';\omega)
  &=&\left[\hat V_{\rm p-h}^{(\rm LS)}(\qvec)*
    \chi_0*\hat V_{\rm p-h}^{(\rm LS)}(\qvec)\right]\nonumber\\
  &=&\frac{1}{N}\left[\VLSq(q)\right]^2\Tr_{\bsigma''}\sum_{\hvec''}
  \left[\I \hat\qvec\times(\hvecF-\hvecF'')\cdot\frac{1}{2}(\bsigma+\bsigma'')\right]\times\nonumber\\&&\times\chi_0(\qvec,\hvec'';\omega)\left[\I
  \hat\qvec\times(\hvecF''-\hvecF')\cdot\frac{1}{2}(\bsigma''+\bsigma')\right]\,.
\end{eqnarray}
All terms with a single $\bsigma''$ operator vanish because
$\Tr_{\bsigma''} \bsigma'' = 0$.  Also, all first-order terms in $(\hat
\qvec\times\hvecF'')$ vanishes because of azimuthal symmetry $
\sum_{\hvec''}(\hat \qvec\times\hvecF'')\chi_0(q,\hvec'';\omega)=0\,.$
Thus, we are left with
\begin{eqnarray}
  \hat W_{\rm LS}^{(2)}(\qvec,\hvec,\hvec',\bsigma,\bsigma';\omega)
  &=&\frac{\left[\tilde V_{\rm p-h}^{\LSsup}(q)\right]^2}{4N} \sum_{\hvec''}
  \chi_0(\qvec,\hvec'';\omega)\times\nonumber\\
  &\times&\Tr_{\bsigma''} \Biggl[\left[(\hat\qvec\times\hvecF)\cdot\bsigma\right]
    \left[(\hat\qvec\times\hvecF')\cdot\bsigma'\right]\nonumber
    +\left[(\hat\qvec\times\hvecF'')\cdot\bsigma\right]
    \left[(\hat\qvec\times\hvecF'')\cdot\bsigma'\right]
    \nonumber\\ &&
    \quad+\left[(\hat\qvec\times\hvecF)\cdot\bsigma''\right]
    \left[(\hat\qvec\times\hvecF')\cdot\bsigma''\right]
    +\left[(\hat\qvec\times\hvecF'')\cdot\bsigma''\right]
    \left[(\hat\qvec\times\hvecF'')\cdot\bsigma''\right]\Biggr]\,.
\end{eqnarray}
Carrying out the $\sigma''$ summation gives the final result
\begin{eqnarray}
  \hat W_{\rm LS}^{(2)}(\qvec,\hvec,\hvec',\bsigma,\bsigma';\omega) &=& 
  \frac{1}{4}\left[\tilde V_{\rm p-h}^{\LSsup}(q)\right]^2\Biggl[\chi_0^{(\perp)}(q;\omega)
    \left[\1+\frac{1}{2}\hat T\right]+\chi_0(q;\omega)\left[\hat Q_7 + \frac{1}{2}\hat Q_9\right]
    \Biggr]\nonumber\\
  &\equiv&\sum_{\alpha,\rm odd} \tilde W_{\rm LS}^{(2,\alpha)}(q;\omega)\hat Q_\alpha
  \label{eq:VLSVLS2}
\end{eqnarray}
where we have introduced the two additional operators
\begin{subequations}
\begin{eqnarray}
  \hat Q_7 &\equiv& \left[(\hat\qvec\times\hvecF)\cdot\bsigma\right]
  \left[(\hat\qvec\times\hvecF')\cdot\bsigma'\right]\,,\\
  \hat Q_9 &\equiv& 2(\hat\qvec\times\hvecF)\cdot(\hat\qvec\times\hvecF')\,,
\end{eqnarray}
\end{subequations}
see Eqs. \eqref{eq:Q79def}. From the orthogonality relations
\eqref{eq:Qortho} we can conclude that the sum of all chain diagrams
containing an even number of spin-orbit interaction operators $\hat
V_{\rm p-h}^{\LSsup}(\qvec)$ {\em and no other   interaction component\/} can be written as
\begin{equation}
  \hat W^{(\rm even)}_{\rm LS}(\qvec,\hvec,\hvec',\bsigma,\bsigma';\omega) =
  \sum_\alpha \tilde W_{\rm LS}^{({\rm even},\alpha)}(q;\omega)\hat Q_\alpha
  \label{eq:LSeven}
\end{equation}
with
\begin{subequations}
\begin{eqnarray}
  \tilde W_{\rm LS}^{({\rm even},c)}(q;\omega)&=&\frac{1}{4}\frac{\left[\VLSq(q)\right]^2
    \chi_0^{(\perp)}(q;\omega)}{
    1-\frac{1}{4}\chi_0(q;\omega)\chi_0^{(\perp)}(q;\omega)\left[\VLSq(q)\right]^2}\,,\\
  \tilde W_{\rm LS}^{({\rm even},T)}(q;\omega)&=&\frac{1}{8}\frac{\left[\VLSq(q)\right]^2
    \chi_0^{(\perp)}(q;\omega)}{
    1-\frac{1}{8}\chi_0(q;\omega)\chi_0^{(\perp)}(q;\omega)
    \left[\VLSq(q)\right]^2}\,,\\
  \tilde W_{\rm LS}^{({\rm even},7)}(q;\omega)&=&
  \frac{1}{4}\frac{\left[\VLSq(q)\right]^2\chi_0(q;\omega)}
       {1-\frac{1}{4}\chi_0(q;\omega)\chi_0^{(\perp)}(q;\omega)
         \left[\VLSq(q)\right]^2}\,,\\
  \tilde W_{\rm LS}^{({\rm even},9)}(q;\omega)&=&
  \frac{1}{8}\frac{\left[\VLSq(q)\right]^2\chi_0(q;\omega)}{1-
    \frac{1}{8}\chi_0(q;\omega)\chi_0^{(\perp)}(q;\omega)\left[\VLSq(q)\right]^2}\,.
\end{eqnarray}
\end{subequations}
There is no contribution to the longitudinal channel $\tilde W_{\rm
  LS}^{({\rm even},L)}(q;\omega)$.\\

To obtain the odd-order chains, one needs to introduce one more new operator
\begin{equation}
  \LSp \equiv \frac{\I}{2}\left[(\hat\qvec\times\hvecF)\cdot\bsigma -
    (\hat\qvec\times\hvecF')\cdot\bsigma'\right]\,.
\end{equation}
For the further manipulations we need the following convolution properties: 
\begin{subequations}
  \label{eq:chains}
    \begin{eqnarray}
    \left[\1* \chi_0*\LS\right] &=& -\frac{\I}{2}\chi_0(q;\omega)
    (\hat\qvec\times\hvecF')\cdot\bsigma'\,,\\
    \left[\hat T*\chi_0*\LS\right] &=& -\frac{\I}{2}\chi_0(q;\omega)
    (\hat\qvec\times\hvecF')\cdot\bsigma\,,\\
    \left[\LSp
      *\chi_0*\LS\right] &=&\phantom{-} \frac{1}{4}[\chi_0(q;\omega)\hat Q_7
      +\chi_0^{(\perp)}(q;\omega)\1]\,,\\
    \left[\hat Q_7*\chi_0*\LS\right]&=&\phantom{-} 
    \frac{\I}{2}\chi_0^{(\perp)}(q;\omega)(\hat\qvec\times\hvecF)\cdot\bsigma\,,\\
    \left[\hat Q_9*\chi_0*\LS\right]&=&\phantom{-} 
    \I\chi_0^{(\perp)}(q;\omega)(\hat\qvec\times\hvecF)\cdot\bsigma'\,,\\
    \left[\1*\chi_0*\LSp\right] &=&-\frac{\I}{2}\chi_0(q;\omega)
      (\hat\qvec\times\hvecF')\cdot\bsigma'\,,\\
    \left[\hat T*\chi_0*\LSp\right] &=&\phantom{-} 0\,.%\\
    %\left[\LS*\chi_0*\hat T*\chi_0*\LS\right] &=&\phantom{-} \chi_0^2(q;\omega)
    %(\hat\qvec\times\hvecF)\cdot(\hat\qvec\times\hvecF')\,,\\
    %\left[\LS*\chi_0*\1*\chi_0*\LS\right] &=&\phantom{-} \chi_0^2(q;\omega)
    %\hat Q_7\,,\\
    %\left[\LSp* \chi_0*\1*\chi_0*\LSp\right] &=&\phantom{-} \chi_0^2(q;\omega)
    %\hat Q_7\,,\\
    %\left[\LSp*\chi_0*\1*\chi_0*\LS\right] &=&\phantom{-} \chi_0^2(q;\omega)
    %\hat Q_7\,.
  \end{eqnarray}
\end{subequations}
The reversed order of operators, \ie, $[\hat O_i * \chi_0 * \hat
  O_j]\leftrightarrow[\hat O_j * \chi_0 * \hat O_i]$, is obtained by
exchanging $\hvecF\leftarrow\hvecF'$, $\bsigma\leftrightarrow\bsigma'$
and $\I\leftrightarrow -\I$. From that, we obtain that the sum of all
odd-power pure $\hat V_{\rm p-h}^{\LSsup}(\qvec)$ chains is a linear combination of the operators
$\LS$ and $\LSp$ in the form
\begin{eqnarray}
  \hat W_{\rm LS}^{(\rm odd)}(\qvec,\hvec,\hvec',\bsigma,\bsigma';\omega) &=&
  \frac{\VLSq(q)}{1-\frac{1}{8}\chi_0(q;\omega)\chi_0^{(\perp)}(q;\omega)\left[\VLSq(q)\right]^2}\left[\LS - \frac{1}{2}\LSp\right]\nonumber\\
  &+&\frac{1}{2}\frac{\VLSq(q)}{1-\frac{1}{4}\chi_0(q;\omega)\chi_0^{(\perp)}(q;\omega)\left[\VLSq(q)\right]^2}\LSp\nonumber\\
  &\equiv&\tilde W_{\rm LS}^{\LSsup}(q;\omega)\LS  +
  \tilde W_{\rm LS}^{\LSpup}(q;\omega)\LSp\,.
 \label{eq:VLSoddApp}
\end{eqnarray}

\subsection{Chain-diagram summation for the full operator structure}
\label{sssec:ChainCLTLS}
An important relationship to obtain the chain-diagram summation of all
operators is
\begin{equation}
  \biggl[\{\1,\hat T\}* \chi_0*\{\LS,\LSp\}* \chi_0*\{\1,\hat T\}\biggr]=0\,,
  \end{equation}
from which we conclude
\begin{equation}
  \biggl[\{\1,\hat T\}*\chi_0*\hat W^{(\rm odd)}_{\rm LS}* \chi_0
      *\{\1,\hat T\}\biggr]=0\,. \label{eq:midodd}
\end{equation}
That is, no terms with an odd number of $\hat V_{\rm p-h}^{\LSsup}(\qvec)$
operators can exist between $\1$ and $\hat T$. Chains that
are combinations of $\hat V_{\rm p-h}^{\rm (c)}(q)$ and $\hat V_{\rm
  p-h}^{\rm (T)}(q)$ and all possible even-order $\hat V_{\rm
  p-h}^{\LSsup}(\qvec)$ are then easily summed by using the orthogonality
relations \eqref{eq:Qortho}.

For a compact representation, redefine the particle-hole interactions
\begin{subequations}
\begin{eqnarray}
  \tilde V_{\rm p-h}^{\rm (c)}(q)&\rightarrow&\tilde V_{\rm p-h}^{\rm (c)}(q;\omega)\equiv \tilde V_{\rm p-h}^{\rm (c)}(q)
  +\frac{1}{4}\chi_0^{(\perp)}(q;\omega)\left[\VLSq(q)\right]^2\,,\label{eq:VcredefApp}\\
  \tilde V_{\rm p-h}^{\rm (T)}(q)&\rightarrow&\tilde V_{\rm p-h}^{\rm (T)}(q;\omega)\equiv\tilde V_{\rm p-h}^{\rm (T)}(q)+
  \frac{1}{8}\chi_0^{(\perp)}(q;\omega)\left[\VLSq(q)\right]^2\label{eq:VTredefApp}\,,
\end{eqnarray}
\end{subequations}
which are now energy dependent. The longitudinal interaction, $\tilde
V_{\rm p-h}^{\rm (L)}(q)$, is unchanged.  That way, the contribution of
all terms containing only sub-chains of an even number of spin-orbit
operators in the $\hat Q_\alpha\in\{\1, \hat L, \hat T\}$ channels can be
written as
\begin{equation}
  \hat W^{\rm even}(\qvec,\hvec,\hvec',\bsigma,\bsigma';\omega) =
  \sum_\alpha \tilde W^{({\rm even},\alpha)}(q;\omega)\hat Q_\alpha
\end{equation}
where all channels of $\tilde W^{({\rm even},\alpha)}(q;\omega)$ are
given by Eqs. \eqref{eq:decp} with $\tilde V_{\rm p-h}^{(\alpha)}(q)$
replaced by $\tilde V_{\rm p-h}^{(\alpha)}(q;\omega)$ defined in Eqs.
\eqref{eq:Vcredef} and \eqref{eq:VTredef} for $\hat Q_\alpha\in\{\1, \hat
T\}$ channels.  Finally, we need to include those terms in the chains
that contain odd-order chains of $\hat V_{\rm p-h}^{\LSsup}(\qvec)$, as
well as $\hat Q_7$ and $\hat Q_9$ channels in the
\eqref{eq:LSeven}. The total effective interaction then has the form
\begin{eqnarray}
  \hat W &=& \hat W^{(\rm even)} + \hat W_{\rm LS}^{(\rm odd)} +\sum_{\alpha\in\{\rm c,T\}}\biggl[\tilde V_{\rm p-h}^{(\alpha)}
  +\tilde V_{\rm p-h}^{(\alpha)}\chi_0\tilde W^{({\rm even},\alpha)}\biggr]
  \hat W^{(1\,{\rm odd},\alpha)}\nonumber\\
  &+& \sum_{\alpha\in\{\rm c,T\}}\biggl[\tilde V_{\rm p-h}^{(\alpha)}+\left[\tilde V_{\rm p-h}^{(\alpha)}\right]^2\chi_0
  +\left[\tilde V_{\rm p-h}^{(\alpha)}\right]^2\chi_0^2\tilde W^{({\rm even},\alpha)}\biggr] \hat W^{(2\,{\rm odd},\alpha)}
\end{eqnarray}
where
\begin{equation}
  \hat W^{(1\,{\rm odd},\alpha)} = \
  \left[\hat W^{(\rm odd)}*\chi_0*\hat Q_\alpha
    + \hat Q_\alpha*\chi_0*\hat W^{(\rm odd)}\right]
\end{equation}
represents chaining with $\hat W^{(\rm odd)}$ on either side, and 
\begin{equation}
  \hat W^{(2\,{\rm odd},\alpha)} = \left[\hat W^{(\rm odd)}*\chi_0*\hat Q_\alpha*\chi_0*\hat W^{(\rm odd)}\right]
\end{equation}
represents chaining with $\hat W^{(\rm odd)}$ on both sides.

Working out these relationships using Eqs. \eqref{eq:chains} we end up with
the compact representation of the remaining contributions to the
effective interaction:
\begin{subequations}
  \label{eq:spinorbitchains}
  \begin{eqnarray}
    \tilde W^{(7)}(q;\omega) &=&\frac{1}{4}\frac{\left[\VLSq(q)\right]^2\chi_0(q;\omega)}{1-\chi_0(q;\omega)\tilde V_{\rm p-h}^{\rm (c)}(q;\omega)}\,,\\
    \tilde W^{(9)}(q;\omega) &=&\frac{1}{8}\frac{\left[\VLSq(q)\right]^2\chi_0(q;\omega)}
   {1-\chi_0(q;\omega)\tilde V_{\rm p-h}^{\rm (T)}(q;\omega)}\,,\\
    \tilde W^{\LSsup}(q;\omega) &=&\frac{\VLSq(q)}{1-\chi_0(q;\omega)\tilde V_{\rm p-h}^{\rm (T)}(q;\omega)}\,,\\
    \tilde W^{\LSpup}(q;\omega) &=&
    \frac{\VLSq(q)\chi_0(q;\omega)V_{\rm p-h}^{\rm (c)}(q;\omega)}
      {1-\chi_0(q;\omega)\tilde V_{\rm p-h}^{\rm (c)}(q;\omega)}%\nonumber\\
      -\frac{\VLSq(q)\chi_0(q;\omega)V_{\rm p-h}^{\rm (T)}(q;\omega)}
      {1-\chi_0(q;\omega)\tilde V_{\rm p-h}^{\rm (T)}(q;\omega)}\,.
  \end{eqnarray}
  \end{subequations}

\section{Calculation of $\chi_0^{(\perp)}(q;\omega)$}
\label{app:chi0trans}
In this section, we show details for the calculation of
\begin{equation}
  \chi_0^{(\perp)}(q;\omega)=-\frac{1}{N}\sum_{\hvec}
  n(h)\bar n(\hvec-\qvec)\chi_0(\qvec,\hvec;\omega)
  (\hat{\qvec}\times\hvecF)^2\,.
\end{equation}
We assume in this appendix that all wave numbers are given in units of
the Fermi wave number $\KF$ and all energies in units of the Fermi
energy $\EF$. Also, let $x \equiv \hbar\omega/\EF$.  Note that we use
a slightly different convention as usual \cite{FetterWalecka} in the
sense that the Lindhard function has the dimension of an inverse
energy. The integrals are done in cylindrical coordinates with $\qvec$
pointing in $z$-direction.  To calculate $\chi_0^{(\perp)}(q;\omega)$,
use
\[\left|\hat\qvec\times\hvecF\right|^2
= h^2\sin^2\theta = h^2-z^2 \equiv \rho^2\]
Then we have
\begin{eqnarray}
  \chi_0^{(\perp)}(q;\omega) &=&
  \int\frac{d^3k}{V_{\rm F}} n(k)\bar n(\kvec+\qvec)
 \frac{2 \rho^2 (q^2+2qz)}{(x+\I\eta)^2 - (q^2+2qz)^2}\nonumber\\
 &=& 
 \frac{3}{2}\left[
  \int_{-q/2}^{1} dz\int_0^{\sqrt{1-z^2}}\rho^3 d\rho
  -  \int_{-q/2}^{1-q} dz\int_0^{\sqrt{1-(z+q)^2}}\rho^3 d\rho\right]
 \frac{2(q^2+2qz)}{(x+\I\eta)^2 - (q^2+2qz)^2}\nonumber\\
 &=&\frac{3}{8}\left[
   \int_{-q/2}^{1} dz (1-z^2)^2- \int_{-q/2}^{1-q} dz(1-(z+q)^2)^2\right]
 \times\frac{2(q^2+2qz)}{(x+\I\eta)^2 - (q^2+2qz)^2}\,.
\end{eqnarray}
We get for the real part
\begin{eqnarray}
  \Re \chi_0^{(\perp)}(q;\omega) &=&\frac{3q^2}{32} - \frac{5}{8} +
  \frac{9}{32}\frac{x^2}{q^2}\nonumber\\
&-&\frac{3}{256q^5}\Biggl[
    ((q^2-x)^2-4q^2)^2
    \ln\left|\frac{x-q^2-2q}{x-q^2+2q}\right|
    -((q^2+x)^2-4q^2)^2\ln\left|\frac{x+q^2-2q}{x+q^2+2q}
    \right|\Biggr]\,.
\end{eqnarray}
We obtain for the imaginary part for $q\le 2$
\begin{equation}
  \Im \chi_0^{(\perp)}(q;\omega) = \begin{cases}
    \displaystyle\frac{3\pi x}{32 q^3}\left(q^4-4q^2+x^2\right)
    &\text{for}\quad 0\le x\le 2q-q^2\\
    -\displaystyle
    \frac{3\pi}{256 q^5}\left((q^2-x)^2-4q^2\right)^2&\text{for}\quad
    2q-q^2\le x\le q^2+2q\\
    0 & \text{for}\quad q^2+2q < x
  \end{cases}
\end{equation}
and for $q> 2$
\begin{equation}
  \Im \chi_0^{(\perp)}(q;\omega) = \begin{cases}
    -\displaystyle
    \frac{3\pi}{256 q^5}\left((q^2-x)^2-4q^2\right)^2&\text{for}\quad
    q^2-2q\le x\le q^2+2q\\
    0 & \text{elsewhere}\,.
  \end{cases}
\end{equation}

The frequency integrations in Eqs. \eqref{eq:localization} and
\eqref{eq:ELSring} are best performed by Wick rotation along the
imaginary $\omega$ axis.  For that purpose we need the transverse
Lindhard function on the imaginary $\omega$ axis:
\begin{eqnarray}
  \chi_0^{(\perp)}(q,\I\omega) &=&\frac{3}{32}q^2
-\frac{5}{8}-\frac{9}{32}\frac{x^2}{q^2}\nonumber\\
&+&\frac{3}{32q^3}\left(x^2+q^2(4-q^2)\right)
\arctan\left(4qx,x^2+q^2(q^2-4)\right)\nonumber\\
  &+&\frac{3}{256q^5}\left((x^2+q^2(4-q^2))^2-4x^2 q^4\right)
 \ln\left|\frac{(q-2)^2+x^2}{(q+2)^2+x^2}\right|
\end{eqnarray}
\end{widetext}
    \bibliography{papers}
\bibliographystyle{apsrev4-2}

\end{document}